# Uncovering Temperature-Dependent Exciton-Polariton Relaxation Mechanisms in Perovskites

Madeleine Laitz[1,2], Alexander E. K. Kaplan[3], Jude Deschamps[3], Ulugbek Barotov[3], Andrew H. Proppe[3], Inés García-Benito[4], Anna Osherov[1,2], Giulia Grancini[5], Dane W. deQuilettes[2*], Keith Nelson[3], Moungi Bawendi[3], Vladimir Bulović[1,2*]

[1]Department of Electrical Engineering and Computer Science, Massachusetts Institute of Technology, 77 Massachusetts Avenue, Cambridge, Massachusetts 02139, USA

[2]Research Laboratory of Electronics, Massachusetts Institute of Technology, 77 Massachusetts Avenue, Cambridge, Massachusetts 02139, USA

[3]Department of Chemistry, Massachusetts Institute of Technology, 77 Massachusetts Avenue, Cambridge, Massachusetts 02139, USA

[4]Department of Organic Chemistry, Universidad Complutense de Madrid. Av. Complutense s/n. 28040 Madrid, Spain.

[5]Department of Chemistry & INSTM, University of Pavia, Via Taramelli 14, 27100 Pavia, Italy

*Corresponding Authors

danedeq@mit.edu, bulovic@mit.edu

## Abstract

State-of-the-art hybrid perovskites have demonstrated excellent functionality in photovoltaics and light-emitting applications and have emerged as a promising material candidate for exciton-polariton (polariton) optoelectronics. In the strong light-matter coupling regime, polariton formation and Bose-Einstein condensation (BEC) have been demonstrated at room temperature in several perovskite formulations. Thermodynamically, low-threshold BEC requires efficient scattering to the polariton energy dispersion minimum at $k_{||} = 0$, and many applications demand precise control of polariton interactions. Thus far, the primary mechanisms by which polaritons relax in perovskites remains unclear. In this work, we perform temperature-dependent measurements of polaritons in low-dimensional hybrid perovskite $\lambda/2$ microcavities and demonstrate high light-matter coupling strengths with a Rabi splitting of $\hbar\Omega_{Rabi} = 260 \pm 5$ meV. By embedding the perovskite active layer near the optical field antinode of a wedged microcavity, we are able to tune the Hopfield coefficients by moving the optical excitation along the wedge length and thus decouple the primary polariton relaxation mechanisms in this material for the first time. We observe the thermal activation of a bottleneck regime, and reveal that this effect can be overcome by harnessing intrinsic scattering mechanisms arising from the interplay between the different excitonic species, such as biexciton-assisted polariton relaxation pathways, and isoenergetic intracavity pumping. We demonstrate the dependence of the bottleneck suppression on cavity detuning, and are able to achieve efficient relaxation to $k_{||} = 0$ even at cryogenic temperatures. This new understanding contributes to the design of ultra-low-threshold BEC and condensate control by engineering polariton dispersions concomitant with efficient relaxation pathways, leveraging intrinsic material scattering mechanisms for next-generation polariton optoelectronics.

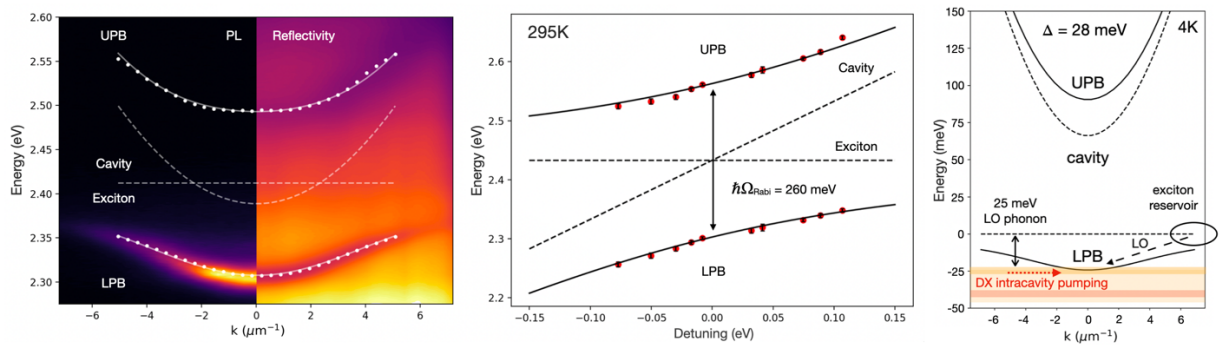

## I. Introduction

Exciton-polaritons (polaritons) are formed in optical microcavities in the strong coupling regime between material excitons and cavity photons. This quantum superposition results in a hybrid state of light and matter, and the formation of a bosonic quasi-particle.[1–3] Polaritons can be modified to adjust the photonic/excitonic character, so that, even when tuned to have a large photonic fraction, polaritons can interact due to the non-zero excitonic component. Such interactions provide several advantages over purely photonic systems in designing logic elements in integrated circuits, such as facile cascadability[4] and large nonlinearities due to the polariton matter component[5], creating the potential for engineering fast, low-power optical transistors.[6,7] These properties also establish opportunities for studying Bose-Einstein condensation[8], quantum vortices[9], and low-threshold polariton lasing[10] for quantum photonic technologies[11] and next-generation qubits[12–16]. Additionally, recent reports utilizing strong coupling to modify electronic structure and energy transfer rates show great promise for the polariton-mediated tuning of chemical reactivity and photophysics.[17] The possibilty of improving energy conversion processes without synthetic changes to the chemical system allows for external modification of kinetics, energy, and electronic and vibrational transitions[18,19] by the precise engineering and control of strongly-coupled devices.[20]

To date, room-temperature polaritons have been sustained in inorganic materials (e.g., GaN, ZnO), organics (e.g., J-aggregates), transition metal dichalcogenides (e.g., $WS_2$, $WSe_2$), and perovskites (e.g., $CsPbBr_3$, $CsPbCl_3$, $(C_6H_5(CH_2)_2NH_3)_2PbI_4$).[21] In perovskites, room-temperature polariton formation[22–28], manipulation[29,30], lasing[31,32], and condensation[5,33–42] have been shown, in which perovskite single crystals, exfoliated flakes, platelets, or thin films are embedded in microcavities typically comprised of either a single distributed Bragg reflector (DBR) and metallic mirror or two high quality DBRs. Although these demonstrations have shown the exceptional potential of perovskite materials for polaritonics, most optoelectronic applications rely on careful control of polariton momentum. Currently, for perovskites, the scattering and relaxation mechanisms that lead to changes in polariton energy and momentum are not well understood. Additionally, there have been very few demonstrations of polariton lasing and condensation in quasi-two-dimensional (herein referred to as 2D) perovskites, which is believed to result from the low room-temperature photoluminescence quantum efficiency (PLQE), low quality factor (Q) cavities leading to short polariton lifetimes, and high exciton-exciton annihilation rates.[21,42–45] Therefore, a deeper understanding of polariton scattering and relaxation mechanisms is needed in order to fully harness polariton utility for a wide range of applications.

Despite these challenges, 2D perovskites remain one of the most promising materials for room-temperature polaritonic devices due to optimal optical properties, chemical versatility, and facile deposition and fabrication schemes. 2D perovskites function as self-assembled quantum well structures, and can be formed as single crystals or polycrystalline thin films. The smaller band gap inorganic monolayers (e.g. $PbI_4$), where the excitons are confined, act as the quantum wells and the larger energy gap organic spacers serve as potential barriers.[46–49] The low Stokes shift, high absorption coefficient, narrow emission, controllable dipole orientation,[28,50,51] and high exciton binding energy (100-500 meV)[52,53] render these materials excellent candidates for facilely fabricated room-temperature polaritonic devices.[21]

Here we explore strong coupling in a test-bed 2D system based on phenethylammonium lead iodide perovskite ($C_6H_5(CH_2)_2NH_3)_2PbI_4$ ($PEA_2PbI_4$) thin films embedded in a wedged microcavity and demonstrate, to the best of our knowledge, a record room-temperature Rabi splitting for $PEA_2PbI_4$ of $\hbar\Omega_{Rabi} = 260 \pm 5$ meV. This figure of merit denotes the magnitude of the photon-exciton coupling,[54] and larger Rabi splittings lead to reduced BEC thresholds as demonstrated by nonequilibrium models[55] and experimental work in organic polaritons.[56] We probe polariton formation by Fourier spectroscopy to image momentum (in-plane $k_{||}$) space via reflectivity and photoluminescence (PL) measurements. We show the emergence of a polariton bottleneck for negative cavity detunings ($E_{cav} - E_{exc} < 0$) and explore temperature-dependent polariton photophysics, revealing, for the first time, the roles of and interplay between reservoir excitons, phonons, and cavity polaritons for extremely efficient polariton relaxation in these materials. We determine the roles of dark excitons and biexcitons in $PEA_2PbI_4$ perovskite polariton scattering mechanisms and intracavity pumping, and the resonant longitudinal optical (LO) phonon interactions mediating efficient relaxation. Additionally, we show that the perovskite electronic structure can be tuned via strong coupling to achieve new excited-state dynamics (i.e. kinetic rates). These insights inform microcavity and polariton dispersion design to harness material-specific, highly efficient polariton relaxation pathways to $k_{||} = 0$ enabling ultra-low-threshold Bose-Einstein condensation in perovskites.[7,57]

## II. Room-Temperature Strong Coupling in 2D Perovskite Microcavities

We realize room-temperature polaritons by fabricating $\lambda/2$ metallic microcavities with a spin-cast $PEA_2PbI_4$ active layer posessing a high degree of crystallinity resembling single crystals (Fig. S1).[58] The strongly-coupled system achieves a Rabi splitting of $\hbar\Omega_{Rabi} = 260$ meV, which is, to the best of our knowledge, the highest reported coupling strength in a $PEA_2PbI_4$ planar microcavity.[21] High-quality epitaxially grown GaAs quantum well microcavities are often fabricated in a wedged geometry, which yields a spatially varying cavity length that allows for the probing of multiple cavity detunings to investigate polaritons with varying photonic/excitonic character.[59,60] By engineering the spin-coating speeds in our solution-processed cavity layers, we achieve a radial wedged cavity, with monotonically increasing cavity length from center to substrate edge which allows for facile changes to the polariton detuning (~30 meV/mm cavity mode gradient, as compared to 13 meV/mm in previously demonstrated epitaxially-grown GaAs quantum well wedged cavities[61]). The reflectivity and photoluminescence dispersions are probed by Fourier spectroscopy, in which the emission of the lower polariton branch (LPB) is only observed in photoluminescence and both the upper polariton branch (UPB) and LPB can be resolved in reflectivity (Fig. 1a-c). To verify minimal change in the polariton coupling strength as a function of position on the cavity, the UPB and LPB energies are extracted from reflectivity measurements and fit at each detuning with Eq. 1,[62] showing that, for each position, the coupling strength remains virtually constant as the cavity mode shifts to lower energies (Fig. 1e, $\hbar\Omega_{Rabi}$ = 260 meV, $\sigma_{Rabi}$ = 5 meV).

$$E_{LP,UP}(k_{\parallel}) = \frac{1}{2}\left[E_{exc} + E_{cav}(k_{\parallel}) \pm \sqrt{4g_0^2 + (E_{exc} - E_{cav}(k_{\parallel}))^2}\right] \quad (1)$$

where $E_{LP,UP}(k_{\parallel})$ corresponds to the upper and lower polariton branch energies as a function of $k_{\parallel}$, $E_{exc}$ is the exciton energy (considered dispersionless over $k_{\parallel}$ values measured), $E_{cav}(k_{\parallel})$ is the uncoupled cavity energy as a function of $k_{\parallel}$, and $2g_0 = \hbar\Omega_{Rabi}$ is the normal mode splitting, or Rabi splitting as in a single-atom microcavity system.[1]

The Hopfield coefficients for each dispersion quantify the excitonic and photonic fraction as a function of $k_{\parallel}$ (Eq. 2), with the most excitonic detuning possessing >50% excitonic character at $k_{\parallel}$ = 0 and the most photonic detuning possessing >50% photonic character at $k_{\parallel}$ = 0 (Fig. 1a-c and Fig. S2, lower panels).

$$|X_k|^2, |C_k|^2 = \frac{1}{2}\left(1 \pm \frac{\Delta E(k_{\parallel})}{\sqrt{\Delta E(k_{\parallel})^2 + 4g_0^2}}\right) \quad (2)$$

where $|X_k|^2$ and $|C_k|^2$ are the excitonic and photonic Hopfield coefficients, respectively.

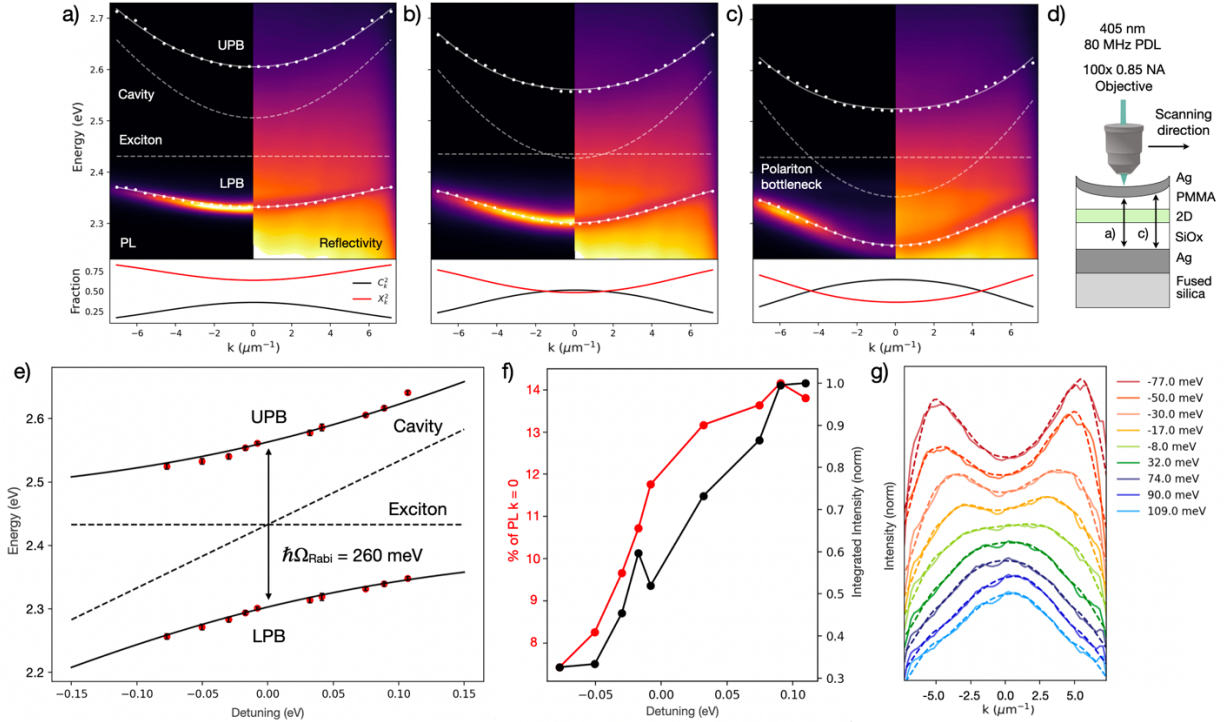

**Figure 1**. Exciton-polariton photoluminescence (PL) (left) and reflectivity (right) dispersions with increasing cavity length from (a) higher cavity mode energy to (c) lower cavity mode energy as shown

schematically in (d). As the cavity shifts to lower energies and the polariton dispersion becomes increasingly photonic (c), the bottleneck effect emerges with the greatest emission intensity at high $k_\parallel$ values. UPB and LPB energies are extracted from reflectivity (white dotted line) and fitted (white solid line) using Eq. 1 (fitted cavity and exciton energies shown, white dashed line) with a Rabi splitting of $\hbar\Omega_{Rabi}$ = 260 meV. (a-c, lower figures) Hopfield coefficients for cavity detunings (photonic fraction $C_k^2$, black trace; excitonic fraction $X_k^2$, red trace) ranging from (a) excitonic to (c) photonic depicting the light-matter characteristics of the generated polaritons as a function of $k_\parallel$ (Eq. 2). (e) Experimental (red dots) and theoretical (black traces) upper and lower polariton branch energies at $k_\parallel$ = 0 with $\hbar\Omega_{Rabi}$ = 260 meV. Dashed black traces correspond to the exciton energy and cavity energy changing with cavity length. (f) As the cavity energy decreases and the dispersion becomes more negatively detuned, the photoluminescence (PL) distribution shifts to higher $k_\parallel$, resulting in a decrease in the fraction of PL within $k_\parallel$ = 0 ± 0.2 μm$^{-1}$ (red trace). Integrated PL intensity as a function of increasingly negative detuning exhibits inhibited emission (black trace). (g) The energy-integrated PL spectra reveal the re-distribution of the maximum PL intensity to higher $k_\parallel$ with increasingly photonic detunings (raw data, solid trace; smoothed data, dashed trace).

As the cavity mode shifts to lower energy and the polariton dispersion becomes more photonically detuned, the emission distribution shifts from the maximum PL intensity at $k_\parallel$ = 0 to higher $k_\parallel$ (Fig. 1f,g), known as the polariton bottleneck regime, a well-studied effect in inorganic and organic polariton systems.[63–70]

The bottleneck effect is a manifestation of the reduced scattering rate of high $k_\parallel$ polaritons as they relax down the LPB. This effect is a combined consequence of the reduced excitonic character as polaritons move down their dispersion curve, which makes exciton-phonon scattering less efficient, and of the reduced density of states as the effective mass decreases, that is, as the LPB curvature increases. In addition, an increase in photonic character is accompanied by an increase in radiative rates, which leads to the depletion of the polariton population near $k_\parallel$ = 0 if the scattering rates into nearby states is too low.[63–65] As the overall excitonic character of polaritons diminishes, and as the LPB curvature increases, the bottleneck effect becomes more pronounced for negative (photonic) detunings. The bulk polariton lifetime at negative detunings is comparatively longer than positive detunings due to the increase in excitonic character of polaritons above the bottleneck region, consistent with our measurements in this system (Fig. S3).

The polariton bottleneck effect has been shown to be thermally activated – by decreasing the system temperature, exciton-phonon scattering is similarly reduced, resulting in the emergence of the bottleneck even for polaritons with mostly excitonic character.[66,67] This phenomenon is difficult to study in systems that require cyrogenic temperatures to achieve strong coupling (e.g., GaAs heterostructures), but room-temperature strong coupling systems present an opportunity to investigate the role of phonon scattering in polariton relaxation by lowering the temperature. Thus, we are able to explore the impact of LO phonon-exciton Fröhlich interactions and acoustic phonon-induced deformation potential mechanisms that can accelerate or hamper key polariton scattering pathways.[67]

### III. Temperature-Dependent Polariton Photophysics in Perovskite Microcavities

By reducing the lattice temperature and carefully controlling the cavity length along the cavity wedge to vary the detuning, we explore the emergence of the bottleneck regime for polaritons with varying excitonic/photonic character and thus changing energy differences between the exciton reservoir and bottom of the lower polariton branch. Additionally, the impact of the magnitude of photon-exciton coupling on temperature-dependent scattering mechanisms is investigated by modifying the Rabi splitting to determine the contribution of enhanced coupling strength on polariton relaxation to $k_{||} = 0$.

To begin deconvoluting these various relaxation mechanisms, we explore the temperature-dependent $k$-space PL distribution for two coupling strengths: $\hbar\Omega_{Rabi} = 175$ meV (Fig. 2) and $\hbar\Omega_{Rabi} = 260$ meV (Fig. S4) by changing the active layer thickness (i.e. number of oscillators $N$, $\hbar\Omega_{Rabi} \propto \sqrt{N}$).[71–73] Additionally, temperature cycles are performed on two detunings ($\Delta = +28$ meV and $+45$ meV). For both positive cavity detunings with emission from $k_{||} = 0$ at 295 K, we probe the emergence of the bottleneck region with decreasing temperature due to the reduction of phonon scattering pathways.

We begin to observe the redistribution of PL to high $k_{||}$ at ~200 K, with the most pronounced bottleneck at ~140 K (Fig. 2). Unexpectedly, the PL distribution in k-space for $\Delta = +28$ meV re-centers at $k_{||} = 0$ for temperatures below 15 K, and, for $\Delta = +45$ meV, re-centers to $k_{||} = 0$ for temperatures below 60 K (Fig. 2d,e). For both detunings, we see the appearance and suppression of the bottleneck effect, with PL moving away from $k_{||} = 0$ when cooling down to 140 K and shifting back to $k_{||} = 0$ when cooling down to 4 K (Fig. 2f, Fig. S4f). To understand this behavior, we note several key factors at play with decreasing temperature, both for the perovskite film system in isolation and for the strongly-coupled microcavity system.

From a material perspective, there are several phenomena that can impact polariton relaxation dynamics intrinsic to the excitonic thin film, including temperature-dependent structural changes, bandgap and concomitant emission energy shifts, and PLQE changes that can alter the polariton density and subsequently the polariton-polariton scattering rates. For the bare 2D film, we confirm that there are no significant changes in the perovskite structure (i.e. phase change) as a function of temperature (Fig. S5), and, as previously shown in PEA$_2$PbI$_4$ 2D perovskites, we observe a redshift in the PL with decreasing temperature which is well captured by the Varshni effect.[74–76] Additionally, we quantify a ~100-fold increase in PLQE, from ~0.7% at 295 K (consistent with other reports)[77] to 77% at 4 K (Fig. S6). While the increase in film PLQE as the temperature decreases serves to increase the polariton population and thus enhance polariton-polariton scattering, this effect is likely not the dominant relaxation mechanism for $k_{||} = 0$ emission at 4 K, as we see very little change to the $k$-space PL distribution as a function of excitation power spanning five orders of magnitude (Fig. S7-S8). If enhanced polariton-polariton scattering due to the increase in material PLQE were the dominant mechanism for relaxation to $k_{||} = 0$ at low temperature, we would expect the re-emergence of the bottleneck effect at sufficiently low powers, or low polariton densities, which we do not observe. Even at low fluences (~30 nJ/cm$^2$/pulse), efficient relaxation to $k_\parallel = 0$ at 4 K is achieved.

For the microcavity system, the exciton PL redshift (~5 nm), combined with the mechanical compression of the microcavity (primarily the optically inert organic spacer layer) resulting in a blue-shift of the cavity energy (~3 nm, Fig. S9), leads to progressive increases in the detuning with decreasing temperature, from $\Delta = +28$ meV at 295 K to $\Delta = +68$ meV at 4 K, or from 70% excitonic character to 80% excitonic character (Fig. 2g). This change in detuning alone is not responsible for the re-centering of the PL distribution around $k_\parallel = 0$, as the bottleneck effect and its suppression are observed even when fixing the detuning by moving to lower energy cavity mode regions at lower temperatures on the radial wedged cavity (Fig. S10).

In the following sections, we will explore the nature of the excitonic species that couple to the cavity mode for low-temperature polariton formation and the mechanisms of bottleneck suppression dependent on both temperature and detuning.

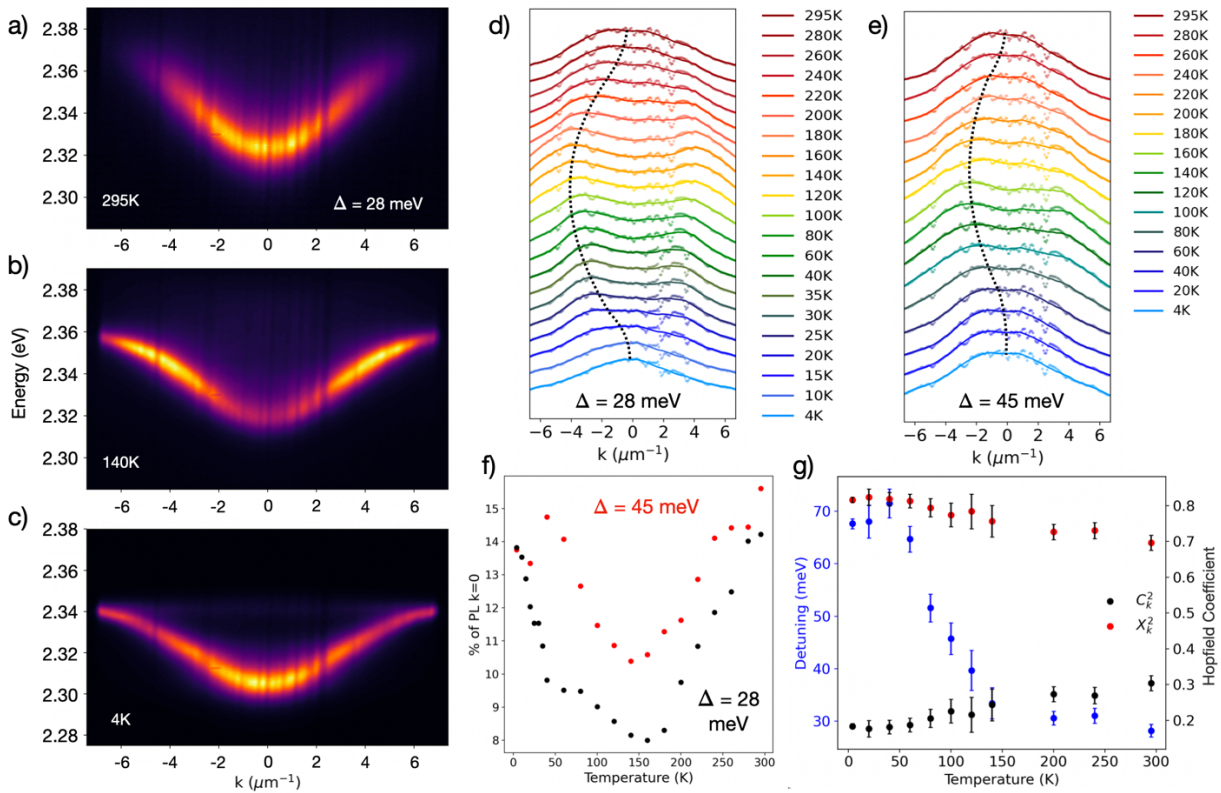

**Figure 2**. (a-c) Lower polariton branch (LPB) photoluminescence (PL) as a function of temperature revealing (d,e) the shifting of the maximum PL intensity to higher $k_\parallel$ at intermediate temperatures before returning to $k_\parallel = 0$ at 4 K. (d-f) The temperature-dependence of the energy-integrated PL was monitored for two detunings established at room temperature ($\hbar\Omega_{Rabi} = 175$ meV, $\Delta = +28$ meV and $+45$ meV), with both detunings showing bottlenecked PL at intermediate temperatures and emission from $k_{||} = 0$ at sufficiently low temperatures (raw data, dashed trace; smoothed data, solid trace; peak PL trend to guide the eye, symmetric about $k_{||} = 0$, dotted trace). (g) Cavity detuning and Hopfield coefficients as a function of temperature for $\Delta = +28$ meV, revealing that the detuning becomes more positive as temperature decreases with polaritons shifting from 70% excitonic at 295K to 80% excitonic at 4 K.

## IV.    2D Perovskite Thin Film Bright-Dark Exciton Spin-Flip Dynamics

To understand the primary mechanism of relaxation at low temperature in this system, a closer look at the uncoupled exciton photophysics (i.e. bare 2D film) is required. In addition to the monotonic PL red-shift of the primary peak (center energy 2.348 eV at 4 K, Fig. S11), a second, low-energy peak emerges below ~140 K, which is consistent with reports of a dark exciton (DX, center energy 2.323 eV at 4 K, Fig. 3a and Fig. S11) shown to increase in PL intensity in the presence of an external magnetic field.[78–80] Emission without a magnetic field has also been observed due to spin-orbit coupling and dipole mixing with the bright exciton (X).[79,80] Though the films posses a high degree of crystallinity (Fig. S5), we are unable to resolve the four states reported within the fine structure for the bright and dark excitons due to PL broadening resulting from disorder of the polycrystalline thin film.[78–83]

Below 80 K, we observe a third, low energy peak below the DX (center energy 2.305 eV at 4 K, Fig. S11), the nature of which has been attributed to electron-phonon coupling, self-trapped excitons, and/or biexcitons (XX) (Fig. 3a).[75,78–80,82,83] The biexciton binding energy is defined as the energy difference between two free excitons and the bound biexciton state, namely $\Delta E_{XX} = 2 \cdot E_X - E_{XX}$. The first stage of biexciton radiative recombination results in an emitted low-energy photon, dictated by the binding energy, ($\hbar\omega_{XX}$) and a remaining free exciton. We attribute this third, low energy peak in the PL to the biexciton, for which we calculate a biexciton binding energy of 43 meV, utilizing PL as a proxy for state energies and relative energies, in agreement with Thouin et al.[84] The assignment to the biexciton species is also consistent with other reports indicating a power-dependent slope of 2 and reporting a short lifetime for the lowest energy PL peak.[79,84–86] The increase in biexciton emission with decreasing temperature is expected with the increase in exciton recombination rates, as radiative recombination begins to outcompete non-radiative Auger-like processes.

Though multiple excitonic species are visible in the low-temperature 2D film PL, single-mode polariton dispersions are observed down to 4 K, direct evidence that the cavity mode couples to only one species.[22,55] To better understand the polariton photophysics and excitonic state to which the cavity couples at low-temperature, we investigate the bare 2D film bright and dark state dynamics by selecting high and low energy regions of the film PL spectrum with spectral filters, capturing the decay of the bright and dark excitons, respectively (Fig. 3). To investigate the dark exciton behavior, these measurements are performed at 80 K and 60 K, temperature regimes in which the dark exciton is present and the biexciton is not the primary emitter.

We observe delayed emission from the dark exciton (low energy spectral region, Fig. 3e,f), which is strong evidence for a spin-flip process,[87–89] which could be facilitated by the dark exciton gaining oscillator strength from the bright exciton by mixing due to spin-orbit coupling.[80,90] This transfer process can be modeled by a set of coupled differential equations. These equations capture the three-state model, in which radiative recombination from the bright exciton (denoted as $X$ in Eq. 3-5) competes with transfer to the dark exciton population (denoted as $DX$ in Eq. 3-5) and vice versa through microscopic reversibility.[91–93] The temperature-dependent spin-flip rate ($k_s(T)$) and the reverse process ($k_{s-}(T)$) are in thermodynamic equilibrium, weighted by the Arrhenius factor: $k_{s-} = e^{(E_X - E_{DX})/kT} \cdot k_s$ (Fig. S12b).[91,94–99]

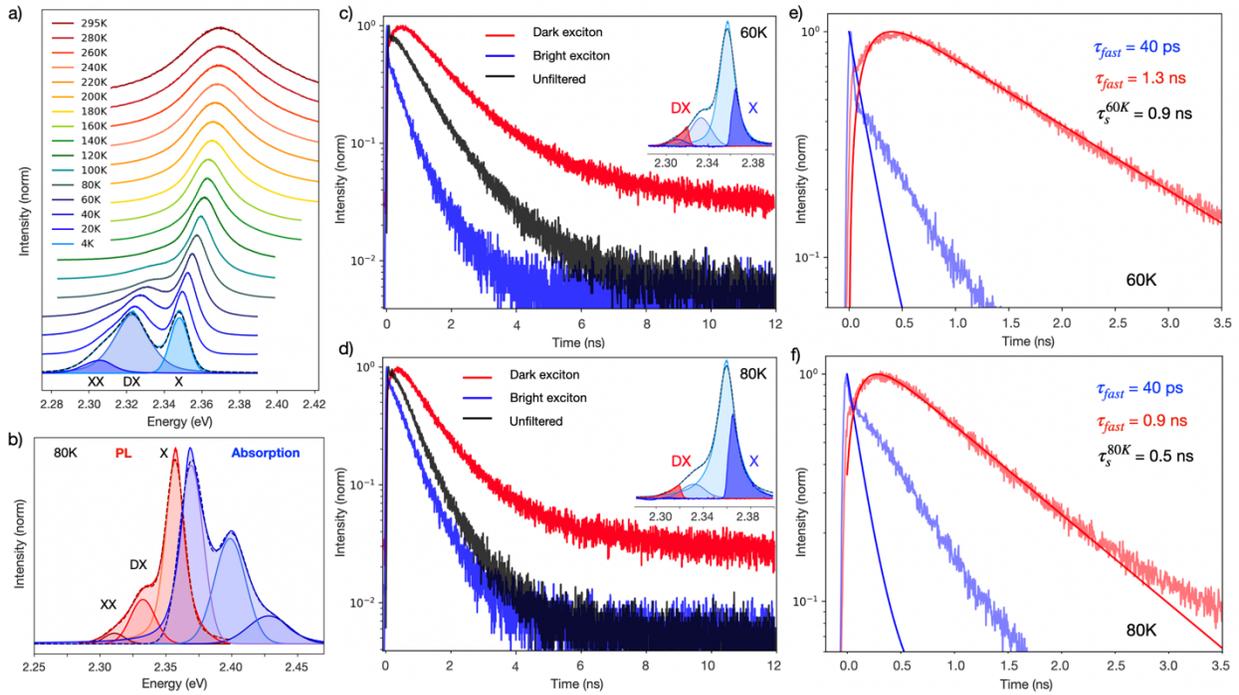

**Figure 3**. (a) 2D perovskite thin film photoluminescence (PL) spectra as a function of temperature revealing, beyond the primary PL peak, the emergence of secondary and tertiary PL peaks assigned to (b) the bright exciton (X), the dark exciton (DX), and the biexciton (XX) with PL shown in red and absorption in blue. (c,d) Time-resolved PL for the unfiltered bare 2D film spectrum (black), X filtered spectrum (blue), and DX filtered spectrum (red). Insets: 2D perovskite thin film PL at (c) 60K and (d) 80K with fitted Gaussian constituent peaks. High (dark blue) and low (dark red) energy regions highlighted, showing the experimentally measured PL spectra with tunable edge-pass filters for X and DX decay measurements, respectively. (e,f) Lifetimes of the X emission (blue) and DX emission (red) simulated with Eqs. 3-5.

To fully capture the exciton/photon dynamics in the photoexcited 2D perovskite film, we take into account photon recycling due to the low material Stokes shift and high absorption (Fig. 3b), creating additional terms that contribute to the excited state dynamics through radiative recombination of either a bright or dark exciton (Eq.3, 5). Photon recycling occurs when a photon generated via radiative recombination is re-absorbed by the active layer.[100] These multiple absorption/emission events serve to increase the carrier density in the thin film for a given excitation power as compared to a material that does not exhibit photon recycling. The impact of photon recycling is dependent on the material absorption coefficient, index of refraction, radiative lifetime, and PLQE (SI).[100,101] At low temperature, the perovskite material PLQE eventually reaches 100-fold the room temperature PLQE, indicating that photon recycling plays an ever-greater role with decreasing temperature.[101]

$$\frac{dn_X}{dt} = -k_{rX}n_X - k_s(T)n_X + k_{s-}(T)n_{DX} + \frac{c}{n_r}\sum_\lambda \alpha_\lambda \gamma_\lambda \qquad (3)$$

$$\frac{dn_{DX}}{dt} = -k_{rDX}n_{DX} + k_s(T)n_{DX} - k_{s-}(T)n_{DX} \qquad (4)$$

$$\frac{d\gamma_\lambda}{dt} = -\frac{c}{n_r}\sum_\lambda \alpha_\lambda \gamma_\lambda + (1 - P_{esc}) \cdot [k_{rX}n_X + k_{rDX}n_{DX}] \qquad (5)$$

where $k_{rX}$ and $k_{rDX}$ are the radiative recombination of X and DX, respectively, $k_s(T)$, and $k_{s-}(T)$ are the temperature-dependent ($T$) spin-flip rates from X to DX and DX to X allowing for interconversion between both species, respectively, $n_X$ and $n_{DX}$ are the X and DX energy carrier concentrations, respectively, $c$ is the speed of light, $\alpha_\lambda$ is the absorption coefficient averaged over the emission band, $\gamma_\lambda$ is the photon concentration within the film for a given wavelength due to radiative recombination and photon recycling, $n_r$ is the index of refraction, and $P_{esc}$ the probability of a radiatively recombined photon leaving the film within the escape cone.

In this way, $k_s(T)$ can be quantified as a function of temperature in this material system. At 60 K (Fig. 3c,e), the system of coupled differential equations yields a fast bright exciton lifetime $\tau^X_{60K,fast}$ < 40 ps, limited by the instrument response function (IRF) of the detection scheme (Fig. S15). The long tail of the time-resolved photoluminescence (TRPL) trace is not included in the model given system IRF limitations and the non-zero spectral overlap of the dark exciton contributing to the counts at longer timescales. The dark exciton has a fast component, $\tau^{DX}_{60K,fast}$ = 1.3 ns, and a long component, fit with an exponential decay, of $\tau^{DX}_{60K,slow}$ = 15.1 ns (Fig. S16), which is consistent with reports of the long dark exciton lifetime in these materials.[80] The transfer process at 60 K, $\tau^S_{60K}$ = 0.9 ns, is relatively slow[102], but several studies show similarly long timescales for spin dynamics in perovskites and other materials at low temperature.[103–111]

When the temperature is increased from 60 K to 80 K, the lifetime of the bright exciton remains IRF-limited ($\tau^X_{80K,fast}$ < 40 ps), the dark exciton fast-component lifetime decreases ($\tau^{DX}_{80K,fast}$ = 0.9 ns), and the transfer rate increases ($\tau^S_{80K}$ = 0.5 ns), which is anticipated considering the Arrhenius relation and large bright-dark state splitting in this material, and is consistent with a thermally-activated spin-flip (Fig. 3d,f).[91,92,94–98,112,113]

When exciting the film, both the bright and dark states can be populated following photoexcitation, with additional dark state population contributions due to transfer from the bright to the dark state via a thermally-assisted pathway.[114–116] As temperature decreases and non-radiative phonon scattering events are reduced, the long-lived dark state population begins to luminesce. Additionally, the bi-directional interconversion between the bright and dark exciton ($k_s(T)$, $k_{s-}(T)$) has been demonstrated in many material systems,[91,117] and here we quantify the rate of this interconversion as a function of temperature. The spin-flip rate slows and the lower-lying dark state is preferentially populated as temperature decreases, and we observe an increase in the dark exciton PL due to the increased likelihood of emission from this state as $k_{s-}(T)$ slows and radiative recombination outcompetes the prolonged spin-flip back to the bright exciton ($\tau^{s-}_{80K} = 63$ ns; $\tau^{s-}_{60K} = 113$ ns). Thus, we observe a steady increase in the dark exciton emission with decreasing temperature due to both a decrease in non-radiative phonon scattering and slowed spin-flip process, as seen in similar systems.[93] Additionally, the impact of photon recycling on these dynamics is non-trivial, and, for PLQE = 77% at 4 K, the average number of recycling events per emitted photon in the 2D perovskite thin film is 3.5, with recycling events increasing to 13 in the radiative limit (Fig. S13). Photon recycling effectively functions as a mechanism for repopulating the bright state exciton reservoir, allowing for another opportunity of the regenerated bright excitons to spin-flip into the dark state despite relatively slow spin-flip rates.

### V. 2D Perovskite Thin Film Exciton and Strongly-Coupled Cavity LPB Lifetimes

The insights gained from the bare 2D thin film kinetic measurements assist in understanding polariton formation and cavity emission as a function of temperature. We investigate the strongly-coupled cavity LPB decay dynamics as a function of temperature to determine the impact of the thermally-activated X to DX spin-flip on polariton formation and relaxation ($\hbar\Omega_{Rabi} = 175$ meV, $\Delta = +28$ meV, corresponding dispersion curves shown in Fig. S17).

Figure 4 shows the temperature-dependent emission lifetimes for the bare 2D film and strongly-coupled cavity (additional temperature-dependent TRPL traces shown in Fig. S18). At room temperature, the bare 2D film lifetime is $\tau_{295K} = 350$ ps, and the LPB emission lifetime is $\tau_{295K,cav} = 260$ ps, representative of the transient response near the cavity polaritons (Fig. 4a). This decay is different from and longer than the expected polariton lifetime (<100 fs for Q ~ 110), due to the widely observed confounding influence of reservoir states and underlying kinetic processes that are only weakly perturbed by cavity polaritons[20] as well as photoinduced effects unique to the cavity system (metallic and DBR alike). These photoinduced effects include polariton contraction, cavity-modulated excited-state absorption, carrier heating, thermal expansion, time-dependent refractive index changes following photoexcitation, as well as excitation pulse-width limitations – effects that can be further deconvolved through angle-resolved measurements and excitation schemes.[118] For organic systems, it has been shown that the polariton lifetime of low-Q cavities, intrinsically on the order of tens of fs, in practice follows the time-evolution of the fundamental carrier and spin dynamics of material excitons on much longer timescales (ns to μs in duration), resulting in strongly-coupled PL decay dynamics that are similar to bare film exciton dynamics.[20]

In the bare 2D film, as the temperature decreases from 295 K to 100 K, the X emission lifetime decreases and the DX emission emerges with an increasingly long lifetime. The two species are visible as a short-timescale fast component (X) and delayed emission with a long tail (DX), and have been observed by Fang *et al.* (Fig. 4b, dark teal trace, 100 K).[83] Further reductions in temperature show the bright exciton emission contribution increasing at early timescales as its emissive lifetime decreases and $k_s(T)$ slows, with the DX demonstrating an increasingly long emissive lifetime as excitons emitting from the dark state are less likely to undergo the reverse spin-flip process back to the bright state according to $k_{s-}(T)$ (Fig. 4c,d, dark purple and blue traces).

In the strongly-coupled cavity at 100 K, the extent of delayed emission is reduced as compared to the bare 2D film, and the fast component contribution increased due to the strong coupling between the bright exciton and cavity mode to form the short-lifetime polariton emissive state competing with the slowed spin-flip from X to DX (Fig. 4b, light teal trace, 100 K). As the temperature decreases to 60 K in the cavity, the fast emission from the strongly-coupled state begins to outcompete the spin-flip, with a smaller fraction of excitons as compared to the thin film undergoing the spin-flip into the long-lived dark state (Fig. 4c, light purple trace). At 4 K in the cavity, we measure nearly IRF-limited emission largely from the strongly-coupled state and do not observe significant delayed emission or the long tail from the DX emission, likely due to the slow spin-flip rate, $k_s(4K)$ (Fig. 4d, light blue trace).

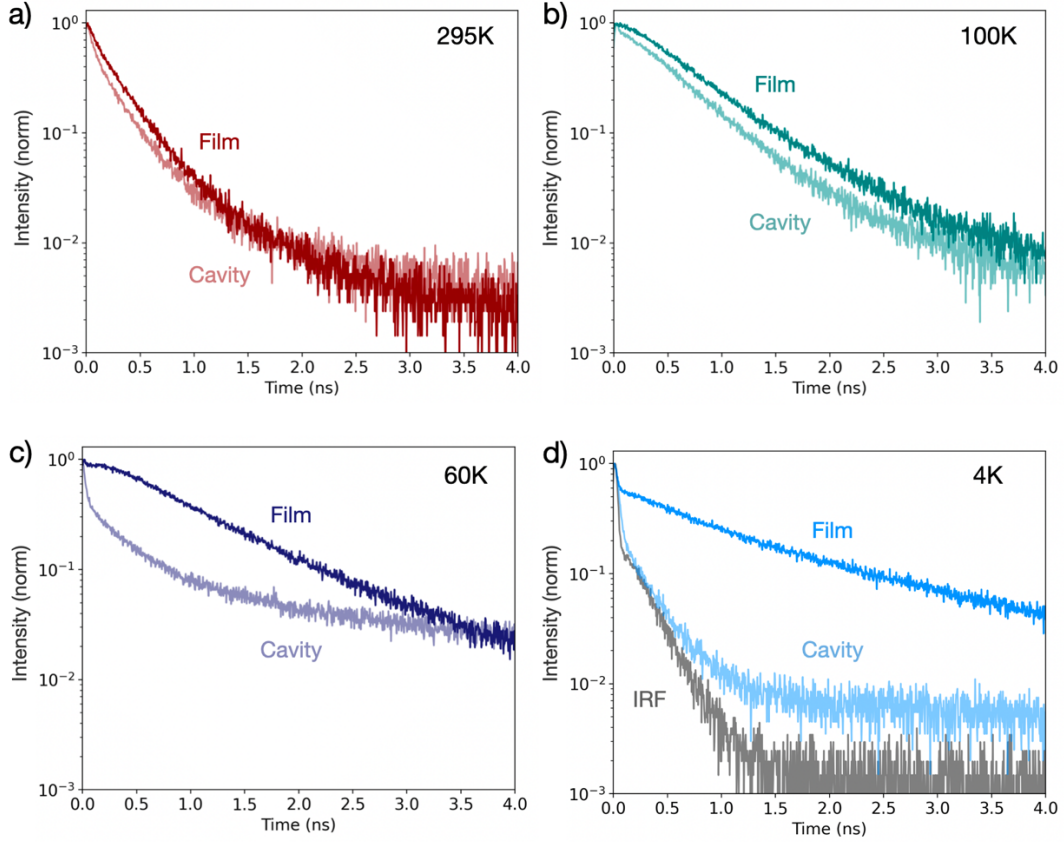

**Figure 4**. (a-d) The bare 2D perovskite film time-resolved photoluminescence (TRPL) decay (dark traces) and the strongly-coupled microcavity ($\hbar\Omega_{Rabi}$ = 175 meV, $\Delta$ = +28 meV) lower polariton branch (LPB) emission (light traces) as a function of temperature (295 K, 100 K, 60 K, 4 K). In the bare film, as temperature decreases, the bright exciton (X) emission lifetime decreases and the dark exciton (DX) emission emerges with an increasingly long lifetime, visible as a short-timescale fast component with delayed emission into a longer tail ((b-c), dark traces). In the cavity ((b-d), light traces), the extent of delayed emission is reduced as compared to the bare 2D film, and the fast component contribution increased due to the additional pathway of the strongly-coupled bright exciton and cavity mode to form the short-lifetime polariton emissive state competing with the spin-flip from X to DX. At 60 K in the cavity ((c), light trace), the fast strong coupling emission competes with the spin-flip and begins to dominate the TRPL decay dynamics at early timescales, with weak emission contribution from the DX state resulting in a long lifetime tail. ((d), light trace) At 4 K in the cavity, a nearly IRF-limited decay is observed with no long lifetime contribution, the strongly-coupled emission likely outcompeting the slow spin-flip rate.

We have shown that, when the perovskite thin film is placed in a microcavity and the bright exciton strongly couples to the cavity photon, a new pathway competes with the spin-flip from the bright exciton to the dark exciton. The bright exciton couples to the cavity and, at temperatures in which the dark exciton begins to luminesce, emits on much shorter timescales than the transfer process, outcompeting the transfer to the dark state, as demonstrated by the low-temperature cavity TRPL traces (Fig. 4c,d). The slow spin-flip rate at low temperature in this material could allow for the kinetic deconvolution of the spin dynamics from the measured polariton lifetime, further approaching the intrinsic polariton lifetime experimentally.

VI. Dark Exciton Intracavity Pumping, Biexciton-Assisted Relaxation, and LO-Phonon-Mediated Bottleneck Suppression

While the strong coupling between the bright exciton and cavity mode results in a short polariton emissive lifetime effectively competing with the slow spin-flip at 4K, there can still exist the dark exciton population generated after photoexcitation.[114–116] This population is similarly likely present in the microcavity system at short timescales (fs to ps),[86,116] though we do not observe its signature long-lived emissive decay in the strongly-coupled LPB low-temperature TRPL trace (Fig. 4d, light blue trace). Similar effects have been demonstrated in strongly coupled systems in which the lower polariton mode exhibits enhanced emissive properties by intracavity pumping from an isoenergetic secondary emissive source within the cavity that itself takes on the luminescence efficiency and kinetic properties of the strongly-coupled state.[119–122] Here, the dark exciton population generated after photo-excitation can benefit from the isoenergetic luminescence pathway created by the strongly-coupled LPB mode, resulting in rapid luminescence decay dictated by the strongly-coupled state. This could lead to increased PL contribution from the dark exciton resonant with the bottom of the LPB in *k*-space (Fig. 5, Fig. S12c).

Additionally, while the cavity does not strongly couple to dark excitons or biexcitons, these species appear resonant with LO-phonon modes in $PEA_2PbI_4$ as computed by Straus *et al.* below 50 meV (400 cm$^{-1}$) using DFT calculations.[75] Straus *et al.* quantify mode contributions from the high energy organic cations (25 meV and 41 meV) resonant with the dark exciton and biexciton energy difference from the bright exciton (X-DX = 25 meV, X-XX = 42 meV).[75,123] In this work, we excite above band gap, and generate hot carriers that can cool via nonradiative relaxation with vibronic replicas – states that resonantly emit phonons into the biexciton state and dark state during spin-flip transfer processes.[124–126] These phonon modes are always present, but it is only when the bottom of the red-shifting LPB is resonant with one of the modes at low temperature that we see LO-phonon assisted relaxation to k∥ = 0 (Fig. 5). When the LPB minimum is resonant with these energies via changing the detuning such that the energy difference between the bottom of the LPB and the exciton reservoir matches the energy of one LO-phonon from either the DX or XX state, efficient LO-phonon-mediated relaxation pathways are utilized.

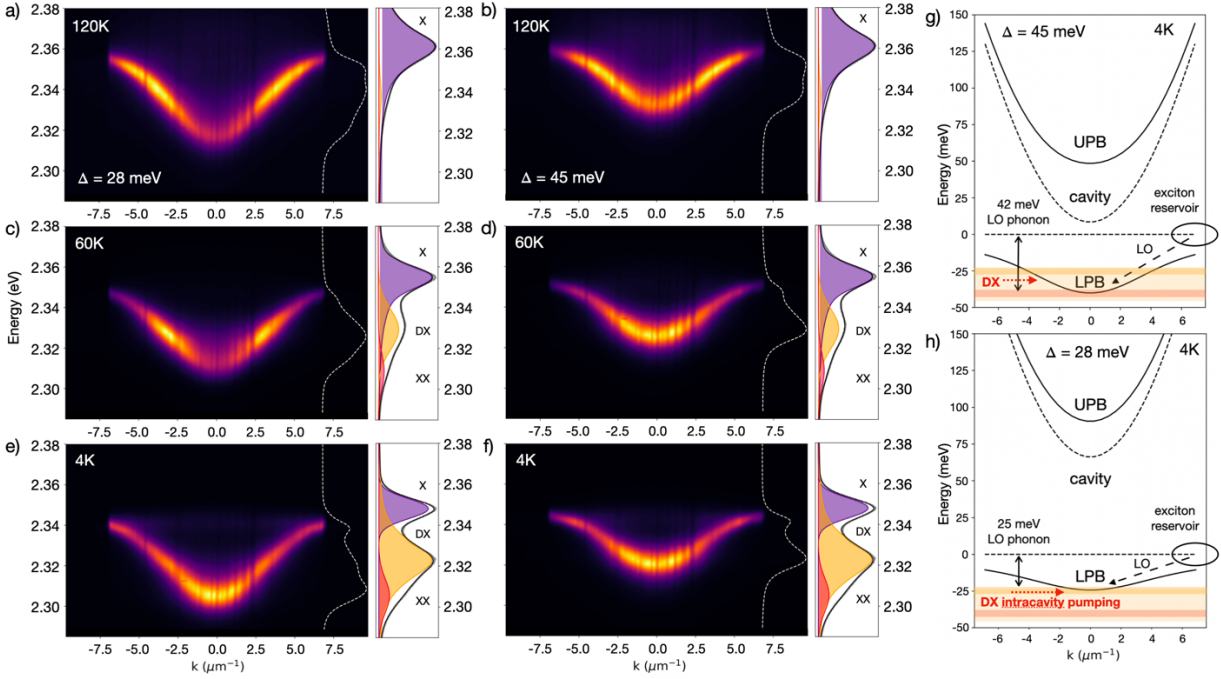

**Figure 5**. (a-f) Lower polariton branch (LPB) photoluminescence (PL) as a function of temperature for two detunings established at room temperature (Δ = +28 meV and +45 meV) with (right) bare 2D perovskite film PL at the corresponding temperature (bright exciton = X; dark exciton = DX; biexciton = XX). (a,c,e) For the more photonic detuning (Δ = +28 meV), the re-centering of the PL distribution to $k_{||}$ = 0 occurs when the bottom of the LPB is resonant with XX, and the polariton benefits from (g) 42 meV LO-phonon-mediated relaxation. ba,d,f) For the more excitonic detuning (Δ = +45 meV), the re-centering of the PL distribution to $k_{||}$ = 0 occurs when the bottom of the LPB is resonant with DX, and the polariton benefits from (h) 25 meV LO-phonon-mediated relaxation.

Emission from $k_{||}$ = 0 at low temperature can be described as a combined consequence of intracavity pumping from the photogenerated dark state population, resonant LO-phonon scattering, and biexciton-assisted relaxation. The less excitonic detuning (Δ = +28 meV) at 120 K (Fig. 5a) shows the brightest emission at high $k_{||}$ (bottleneck), whereas, at 60 K, the brightest emission has shifted to lower $k_{||}$, peaked at the DX energy enhanced by isoenergetic intracavity pumping, and matches the energy of the smaller LO-phonon resonance (Fig. 5c). This resonance assists higher energy polaritons to scatter down the LPB, but it is not until colder temperatures that the bottom of the LPB red-shifts further, becoming resonant with the XX, and then benefits from relaxation via XX LO-phonon pathways for the brightest emission at $k_{||}$ = 0 (Fig. 5e,g).

For the more excitonic detuning (Δ = +45 meV) at 120 K (Fig. 5b), the bottleneck effect is again observed, though to a lesser extent as the increased excitonic character reduces the curvature of the LPB and increases the DOS, allowing for greater scattering towards $k_{||}$ = 0. Because of the flatter dispersion, the bottom of the LPB never reaches the XX energy resonance, but instead is resonant with the DX state and again matches the smaller LO-phonon energy. This occurs at higher temperatures than its less excitonic counterpart, $\Delta = +28$ meV, resulting in efficient $k_{||} = 0$ emission at an elevated temperature of 60 K (Fig. 5d), maintaining this lowest energy, $k_{||} = 0$ primary emission down to 4 K (Fig. 5f,h). These proposed scattering pathways are consistent with reports in inorganic polaritons, identifying LO-phonon assisted transfer – including the first LO-phonon replica of the biexciton – as efficient mechanisms for directly populating $k_{||} = 0$.[57,68,127–130] We can observe additional signatures of biexciton-assisted relaxation given that the biexciton has two viable pathways for dissociation, as described by Corfdir *et al.*[57]:

*Biexciton* → *LPB Polariton at* $k_{||} = 0$ + *LO-Phonon* + *Bright Exciton*

*Biexciton* → *LPB Polariton at* $k_{||} = 0$ + *LO-Phonon* + *LPB Polariton at High* $k_{||}$

In both 4 K dispersions (Fig. 5e,f), there is faint, uncoupled exciton PL in a flat, dispersionless line at the low tail of the bare exciton energy band. This uncoupled PL is only visible for temperatures at which the biexciton emission is prominent (<20 K, Fig. S21). We also observe brightening of the high $k_{||}$ wings of the LPB for T < 15 K (Fig. S21), revealing contributions to the LPB PL distribution from both mechanisms above.

Isoenergetic dark exciton intracavity pumping, LO-phonon-mediated scattering, and biexciton-assisted polariton relaxation are key pathways for $k_{||} = 0$ emission in this system, presenting opportunities for engineering the microcavity detuning such that the bottom of the lower polariton branch can be directly populated by highly efficient non-radiative LO-phonon emission which facilitates targeted polariton scattering. These insights are critical for the rational engineering of microcavity systems enabling low-threshold, optically excited condensation and novel perovskite condensates via electrical injection of carriers one LO-phonon above the bottom of the lower polariton branch.[68,131–134]

## VII. Conclusion

We have identified, for the first time, highly efficient scattering pathways enabled at low temperature in PEA$_2$PbI$_4$ polaritons by changing the cavity detuning to benefit from resonant LO-phonons and biexciton-assisted relaxation, with enhanced emission due to the isoenergetic intracavity pumping from the dark state reservoir. Along with a 100-fold increase in the material PLQE at low temperature, these factors provide further insight into strategies for low-threshold BEC in perovskite strongly-coupled microcavities, in which relaxation efficiency dictates the condensation threshold. Additionally, we explore the exciton fine structure and dynamics, quantifying the spin-flip transfer process from bright to dark excitons at low temperature and investigating the competition between strong coupling and transfer from the bright to dark state. In this way, we demonstrate, without synthetic modifications, tuning of the perovskite electronic structure via strong coupling to enable new kinetic rates. These insights provide cavity and material design principles for next-generation polaritonic devices requiring careful control of polariton momentum and relaxation, and demonstrate the utility of polariton formation to modify recombination dynamics and energy conversion processes towards optoelectronic devices with tunable emissive properties.

**Methods**

*Perovskite Preparation.* Perovskite precursors were obtained from Sigma Aldrich (phenethylammonium iodide, SKU 805904) and TCI (lead(II) iodide, TCI-L0279), and prepared in dimethyl sulfoxide (Sigma Adlrich, SKU 34869) stoichiometrically for n = 1. Films were spincast in a two-step procedure: 1) 1000 rpm, 10 s, 500 accel; 2) 5000 rpm, 30 s, 2000 accel with a chlorobenzene quench 15 s before the end of the second step (Sigma Aldrich, SKU 284513). The films were annealed at 100˚C for 10 min. All synthesis and process steps under nitrogen. The resulting thin films ranged from approximately 25-50 nm (+/- approximately 2 nm) as a function of solution concentration, as measured on silicon substrates by ellipsometry.

*Solution-Processed Spacer Layer Preparation.* Poly(methyl methacrylate) was purchased from Sigma Aldrich (SKU) and dissolved in chlorobenzene (Sigma Aldrich, SKU 284513) at 50˚C. Films were spincast using a single-step procedure: 1500 rpms, 60 s, 1000 accel. Films were gently annealed at 60˚C for 1 min to assist in driving off excess solvent. All process steps under nitrogen. The resulting thin films were approximately 110 nm thick, as measured on fused silica substrates by profilometry.

*Microcavity Preparation.* Fused silica substrates were cleaned by sonication in water, diluted detergent, acetone, and isopropyl alcohol followed by boiling isopropyl alcohol. The bottom Ag mirror was thermally evaporated at 110 nm followed by sputter deposition of a 108 nm SiOx layer in argon. The perovskite active layer was spin-cast and annealed under nitrogen (~25 nm), followed by the Poly(methyl methacrylate) layer. The microcavity was capped with a semi-transparent thermally evaporated Ag layer (35 nm).

*Room-Temperature Fourier Spectroscopy.* K-space was imaged using a Nikon Eclipse-Ti inverted microscope fitted with an infinity corrected 100 × dry objective (Nikon L Plan, NA = 0.85). A 405 nm pulsed diode laser (PDL-800 LDH-P-C-405B, 300 ps pulse width) was used for excitation with repetition rate of 80 MHz. The sample photoluminescence (PL) was filtered through a 405 nm dichroic beamsplitter (Nikon DiO1-R405) and the reflectivity collected via a halogen lamp white light source (Nikon Eclipse-Ti) and 50/50 beamsplitter (Chroma 21014-UF3 C188781). The output for both PL and reflectivity was then coupled in free space via a 4F imaging system into a Princeton Instruments Acton spectrometer and Pixis camera (100 (k-space) x 1340 (wavelength) pixels).

*Low-Temperature (4-295 K) Fourier Spectroscopy.* K-space was imaged using a Montana Instruments closed-cycle liquid He crysotat with piezo-controlled 3D-moveable sample stage, cryo-optic low-working distance 100x 0.9NA objective, vacuum housing, radiation shield, and local objective heater. A wavelength-tunable ultrafast laser (Toptica Photonics FemtoFiber Pro) was used for 488 nm excitation with 80MHz repetition rate, guided into the cryo-optic with electrically-controlled Thorlabs Galvo mirrors. The sample emission was filtered through a Semrock tunable edge pass (set to 490 nm long pass) filter and directed via a 4F imaging system into a Princeton Instruments Acton spectrometer and either a 512 (k-space) x 512 (wavelength) pixel or 1024 (k-space) x 1024 (wavelength) pixel Pixis camera.

*Time-Resolved Photoluminescence.* A wavelength-tunable ultrafast laser (Toptica Photonics FemtoFiber Pro) was used for 488 nm excitation with 80MHz repetition rate. The emission was collimated and filtered through a Semrock tunable edge pass (set to 490 nm long pass) filter and focused onto a Micro Photon Devices (MPD) PicoQuant PDM Series single photon avalanche photodiode with a 50 $\mu$m active area and 40 ps IRF. Photon arrival times were time-tagged using a time-correlated single photon counter (TimeHarp 260).

*Variable Angle Spectroscopic Ellipsometry.* Spectroscopic ellipsometry was performed using a variable angle spectroscopic ellipsometer (Woollam) at 65°, 70°, and 75° angles of incidence. Ellipsometry data was fitted to obtain perovskite thin film thicknesses.

*Photoluminescence Quantum Efficiency (PLQE) Measurements.* PLQE measurements were acquired using a center-mount integrating sphere setup (Labsphere CSTM-QEIS-060-SF) and Ocean Optics USB-4000 spectrometer. The integrating sphere setup was intensity calibrated with a quartz tungsten halogen lamp (Newport 63355) with known spectral irradiance set at a distance 0.5m away from the integrating sphere illumination port. A fiber-coupled 405 nm diode laser in CW mode (PDL-800 LDH-P-C-405B) was collimated with a triplet collimator (Thorlabs TC18FC-405) to produce a beam with an approximate $1/e^2$ diameter of 2.8mm. The beam was used to excite the sample and a variable neutral density filter was used to attenuate the laser. Data acquisition followed the protocol described by de Mello et al.,[135] with a scattering correction.

*Cryo XRD.* Temperature-dependent XRD was performed using a Panalytical Multipurpose Diffractometer with a liquid He cryostat for in-situ low temperature measurements. 30 minute scans were taken for 5-67 degrees at each temperature in increments of 20K from 295K to 11K, with 15 minutes between scans for temperature equilibration. The temperature was scanned from high to low and cycled back from low to high to determine whether the temperature cycle damaged the 2D perovskite. No structural changes (peak intensity or position) were noted in the up cycle.

*Cryo Absorption.* Reflection spectrophotometry was performed with light incident from the film side using an Agilent Cary 5000 dual-beam UV–vis–NIR spectrophotometer with home-built, liquid $N_2$ cryostat quartz window attachment. The 2D perovskite was coated onto a 15 mm diameter fused silica optical window (see substrate cleaning procedure above) for compatibility with cryo sample holder. Specular reflectance was collected at an incident angle of 8°. A 3 mm round aperture was used for all measurements.

**Acknowledgments**

This work is supported by the TATA-MIT GridEdge Solar Research program. This material is based upon work supported by the National Science Foundation Graduate Research Fellowship under Grant No. (1122374). This project has received funding from the European Research Council (ERC) under the European Union's Horizon 2020 research and innovation programme ERC HYNANO (grant agreement n° 802862). M.L. acknowledges support from MIT's Hugh Hampton Young fellowship and Samsung Advanced Institute of Technology. A.E.K.K acknowledges support from the US Department of Energy, Office of Basic Energy Sciences, Division of Materials Sciences and Engineering under award no. DE-SC0021650. J.D. acknowledges support from the NSERC Postgraduate Scholarship program and U.S. Department of Energy. U.B. acknowledges support from the National Science Foundation, Award Number CHE-2108357. A.H.P. acknowledges support from the U.S. Department of Energy, Office of Science, Basic Energy Sciences under Award DE-SC0021650, and the Natural Sciences and Engineering Research Council of Canada (NSERC PDF). We thank Thomas Mahony for valuable discussions and insights; Roberto Brenes for spectrometer software integration and integrating sphere calibration; and Charles Settens for assistance with XRD.

**Contributions**

M.L., A.E.K.K., D.W.D., M.B., and V.B. conceived and designed the experiments. M.L. prepared the films and fabricated wedged cavities with supervision from D.W.D. and support from I.G.B. and G.G.. M.L. designed and performed the optical characterization and analysis of films and cavities at room temperature with support from J.D., K.N., and D.W.D.. The optical characterization of the perovskite films and devices at low temperature was performed by M.L. and A.E.K.K.. The cryo absorption measurements were performed by U.B., and A.O. designed the cryo XRD measurements with support from Charles Settens (*Acknowledgements*). M.L. wrote Python code for analysis and simulations with guidance from D.W.D.. M.L. wrote the first draft of the manuscript with early drafts edited by A.E.K.K. and D.W.D., and all authors contributed feedback and comments. K.N., M.G.B., V.B. and D.W.D. directed and supervised the research.

# Supporting Information

**Uncovering Temperature-Dependent Exciton-Polariton Relaxation Mechanisms in Perovskites**

Madeleine Laitz[1,2], Alexander E. K. Kaplan[3], Jude Deschamps[3], Ulugbek Barotov[3], Andrew H. Proppe[3], Inés García-Benito[4], Anna Osherov[1,2], Giulia Grancini[5], Dane W. deQuilettes[2*], Keith Nelson[3], Moungi Bawendi[3], Vladimir Bulović[1,2*]

[1]Department of Electrical Engineering and Computer Science, Massachusetts Institute of Technology, 77 Massachusetts Avenue, Cambridge, Massachusetts 02139, USA

[2]Research Laboratory of Electronics, Massachusetts Institute of Technology, 77 Massachusetts Avenue, Cambridge, Massachusetts 02139, USA

[3]Department of Chemistry, Massachusetts Institute of Technology, 77 Massachusetts Avenue, Cambridge, Massachusetts 02139, USA

[4]Department of Organic Chemistry, Universidad Complutense de Madrid. Av. Complutense s/n. 28040 Madrid, Spain.

[5]Department of Chemistry & INSTM, University of Pavia, Via Taramelli 14, 27100 Pavia, Italy

*Corresponding Authors**: danedeq@mit.edu, bulovic@mit.edu

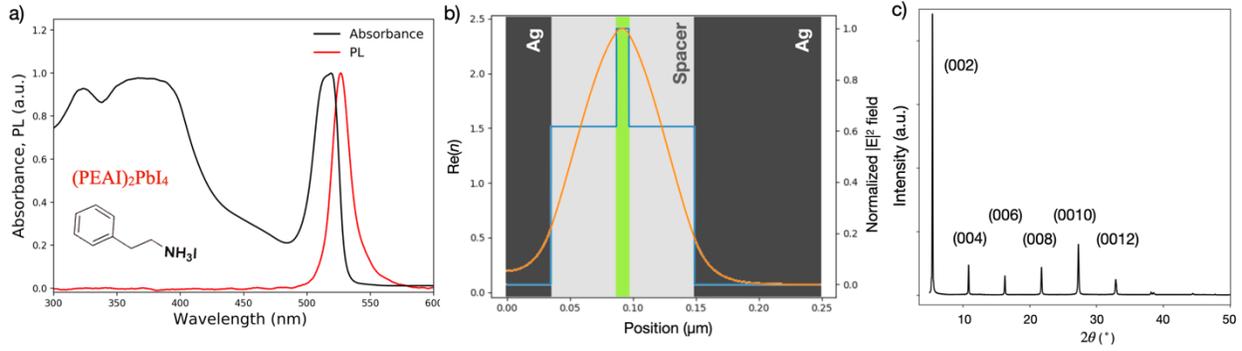

**Figure S1**. (a) Absorption (black trace) and photoluminescence (red trace) spectra for the PEA$_2$PbI$_4$ thin film. (b) Metallic microcavity structure with normalized electric field profile (orange trace) and index of refraction (blue trace) simulated with a transfer matrix model of the cavity architecture: Ag (110nm)/SiOx (108nm)/spin-cast PEA$_2$PbI$_4$ active layer (~20nm)/PMMA (~110nm). (c) Room-temperature XRD demonstrating a high degree of crystallinity equivalent to single crystals.[1]

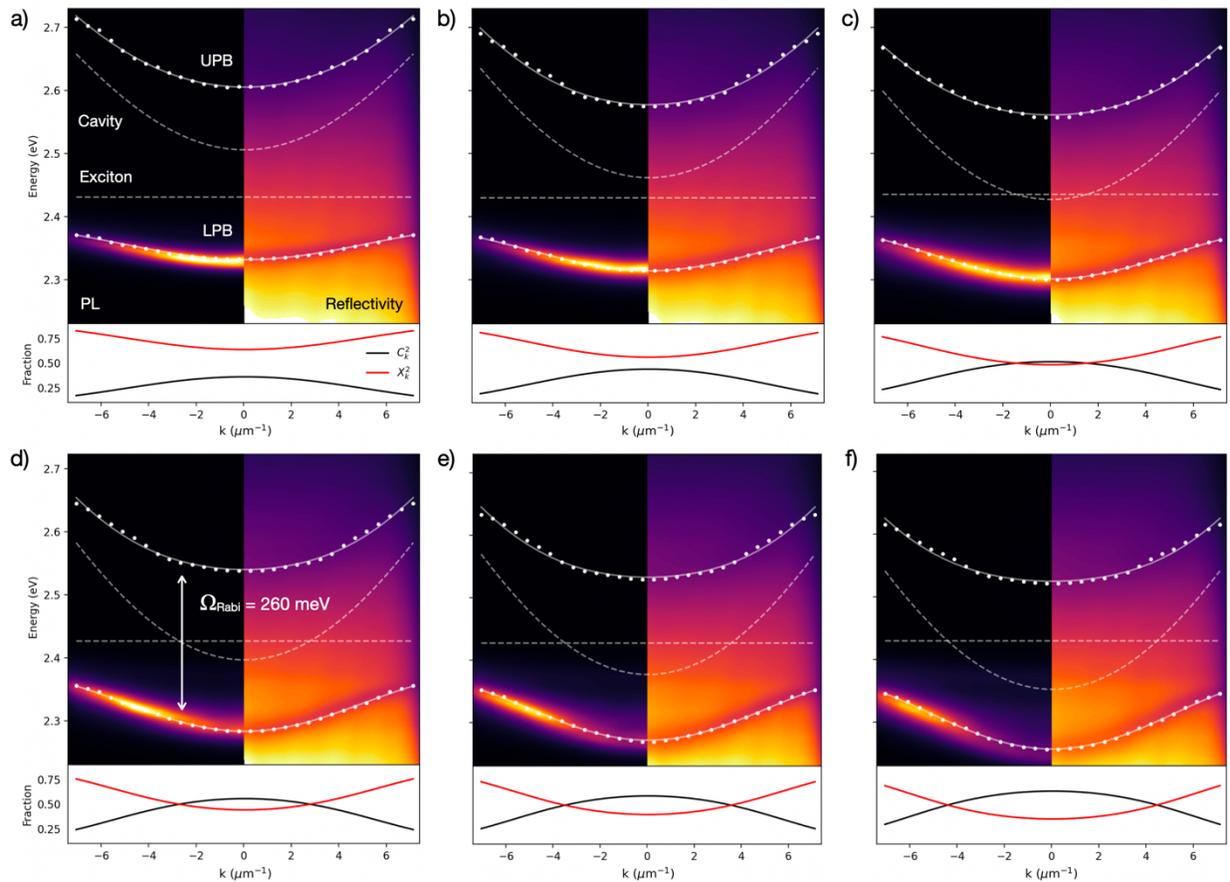

**Figure S2**. Exciton-polariton photoluminescence (left) and reflectivity (right) dispersions with increasing cavity length from (a) higher cavity mode energy to (f) lower cavity mode energy (f).

As the cavity shifts to lower energies and the polariton dispersion becomes increasingly photonic (d-f), the bottleneck effect emerges with the greatest emission intensity at high k values. The upper and lower polariton branches are extracted from reflectivity (white dotted line) and fit (white solid line) with a Rabi splitting of $\hbar\Omega_{Rabi}$ = 260 meV. (a-f, lower figures) Hopfield coefficients for cavity detunings (photonic fraction $C_k^2$, black trace; excitonic fraction $X_k^2$, red trace) ranging from (a) excitonic to (f) photonic depicting the light-matter characteristics of the generated polaritons as a function of $k_{//}$.

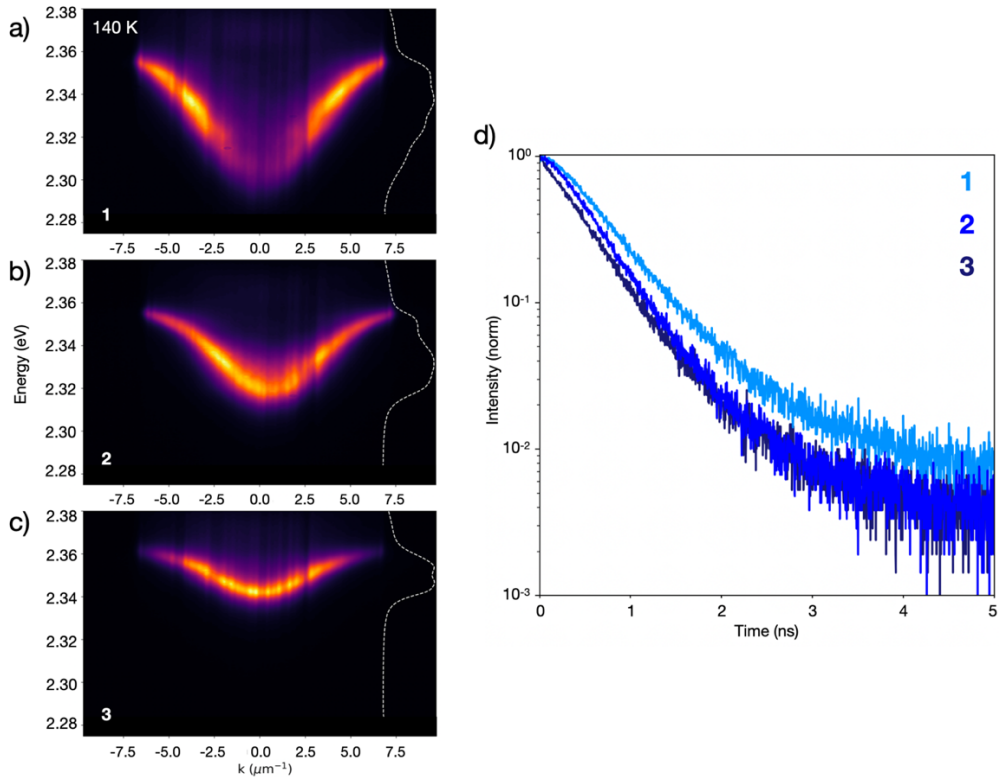

**Figure S3**. Photoluminesence (PL) in k-space for $\hbar\Omega_{Rabi}$ = 175 meV at 140 K for (a) highly photonic (negative) detuning with severe bottleneck, (b) photonic detuning with the beginnings of a bottleneck, and (c) excitonic (positive) detuning with no bottleneck. (d) Increasingly positive detunings result in polaritons with shorter radiative lifetimes due to the suppression of the bottleneck effect (blue traces 1-3 corresponding to (a)-(c), respectively). PL from higher $k_{||}$ in the bottleneck region corresponds to more excitonic polaritons, which possess greater scattering rates and longer radiative lifetimes, accounting for the delayed emission at early timescales and longer lifetime tails.

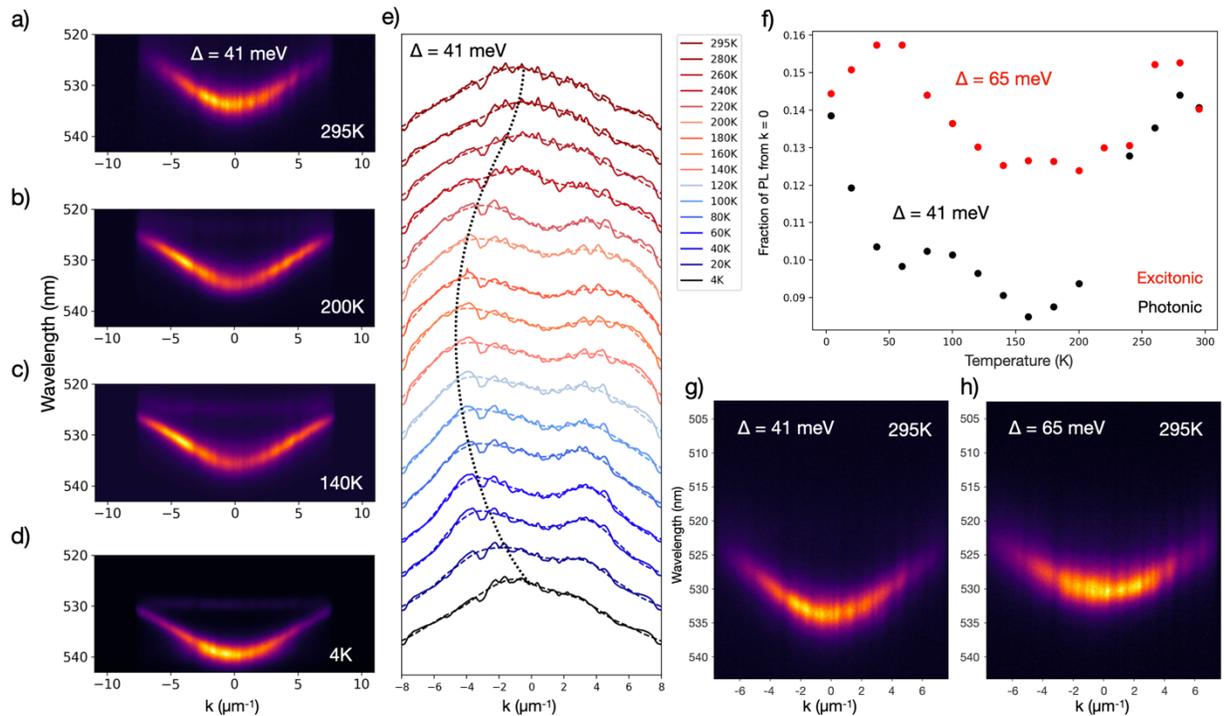

**Figure S4**. (a-d) Lower polariton branch (LPB) photoluminescence (PL) as a function of temperature for $\hbar\Omega_{Rabi}$ = 260 meV revealing e) the migration of the maximum PL intensity to higher $k_{||}$ at intermediate temperatures before returning to $k_{||}$ = 0 at 4K. (e,f) The temperature-dependence of the energy-integrated PL for $\Delta$ = +41 meV showing bottlenecked PL at intermediate temperatures and emission from $k_{||}$ = 0 at sufficiently low temperatures. (g,h) PL spectra for two detunings ($\Delta$ = +41 meV and +65 meV) at 295K, corresponding to the temperature series in (f). Note: asymmetries in PL distribution arise from imperfectly flat substrate seating due to Ag cryo paste used for thermal conductivity. Additionally, the thicker perovskite active layer results in increased uncoupled exciton PL at elevated temperatures as compared to the thinner active layer yielding $\Omega_{Rabi}$ = 175 meV.

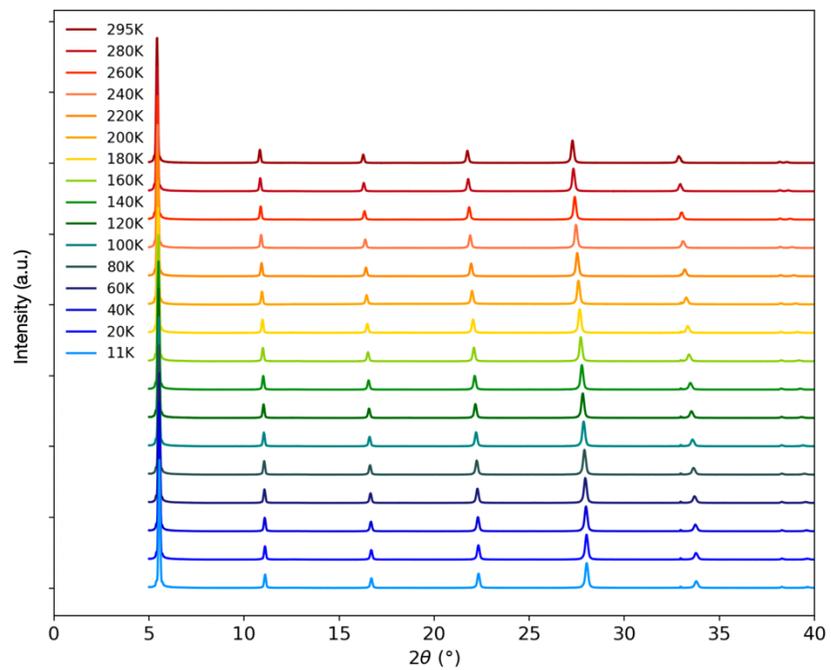

**Figure S5**. Temperature-dependent XRD from 295 K to 11 K showing no phase change as a function of temperature.

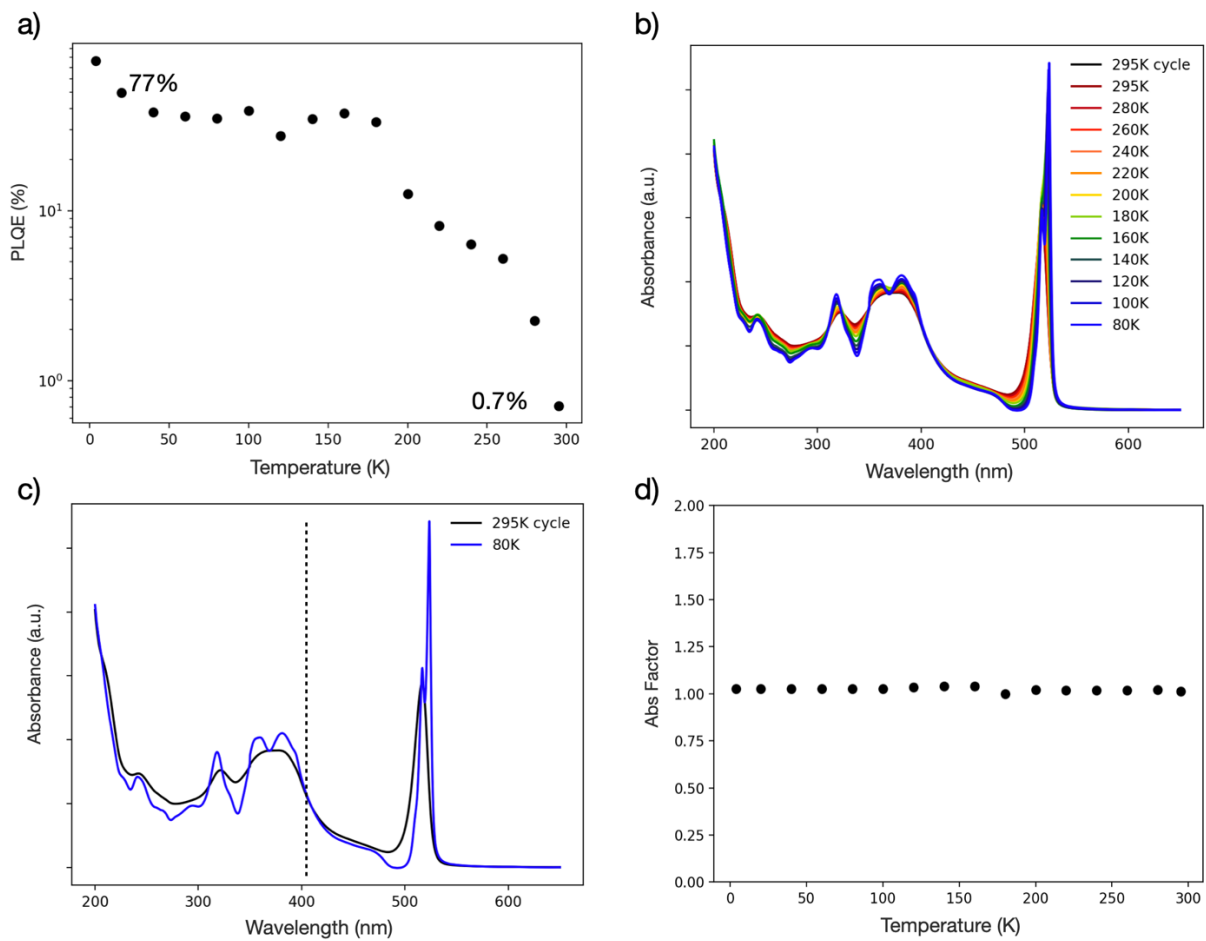

**Figure S6**. (a) PLQE measured at 295 K with integrating sphere and 405 nm laser in cw-mode (~0.7%), and used to calculate a 100-fold PLQE increase as a function of temperature. To ensure the PL increase was not from changes in the absorption of the excitation as a function of temperature, (b) temperature-dependent absorption measurements were performed from 295 K to 80 K. The excitation wavelength was tuned to a region with very little change in absorption (c, dashed black trace indicating laser excitation wavelength), with small fluctuations in absorbance quantified in the (d) Abs Factor extrapolated to 4K.

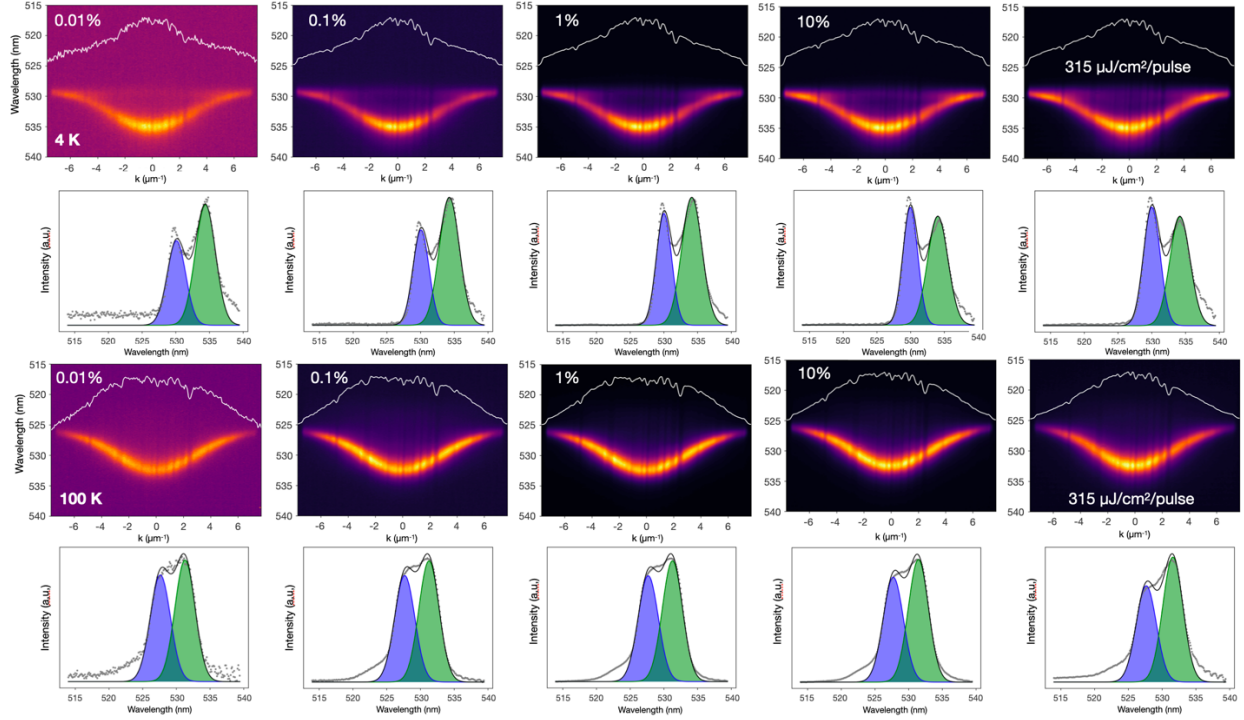

**Figure S7**. Lower polariton branch (LPB) photoluminescence (PL) ($\hbar\Omega_{Rabi}$ = 175 meV, $\Delta$ = +40 meV) as a function of excitation power spanning five orders of magnitude. The top row shows the $k$-space dispersion, and the bottom row shows the $k$-space-integrated PL with high-$k_{||}$ (high energy) and low-$k_{||}$ (low energy) spectral regions fit to determine whether the increase in power results in an increase in emission from the bottom of the LPB due to increased polariton-polariton scattering. No such trends are observed at 4 K, and only a weak increase in the low-$k_{||}$ region is seen at 100 K, indicating that polariton-polariton scattering is likely not the primary factor for the dramatic redistribution of PL to $k_{||}$ = 0 at low temperature. *Row 1,2*: 4 K power series (demonstrates a more rapid increase in high $k_{||}$ PL (blue Gaussian) than low $k_{||}$ (green Gaussian) indicating greater biexciton emission enhancement with increasing power); *Row 3,4:* 100 K power series (ratio between high $k_{||}$ (blue Gaussian) and low $k_{||}$ (green Gaussian) emission preserved).

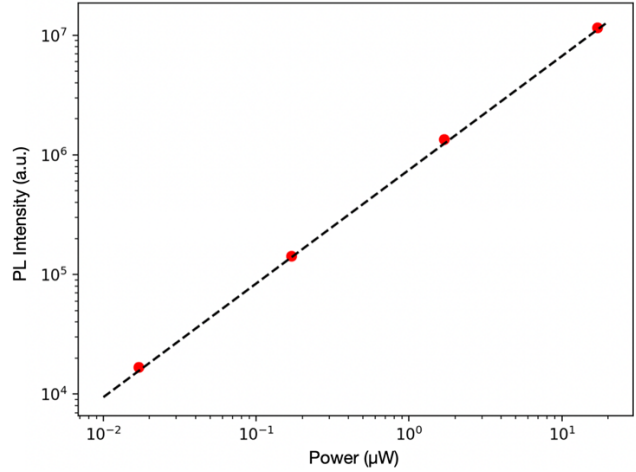

**Figure S8**. Integrated lower polariton branch PL at 4 K for $\hbar\Omega_{Rabi} = 260$ meV spanning four orders of magnitude, revealing a slope of m = 0.95, consistent with bright exciton power dependence.[2–5]

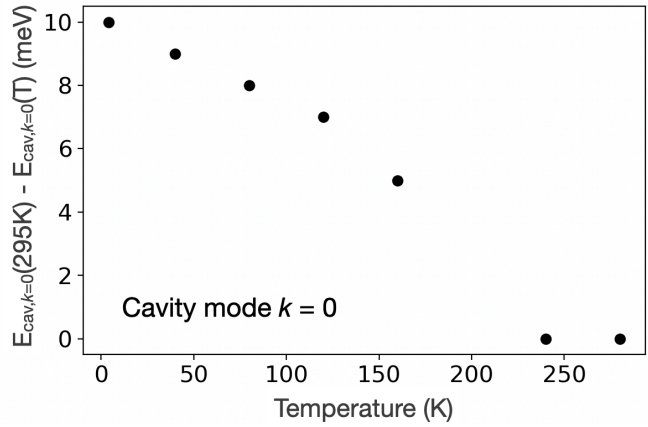

**Figure S9**. Cavity mode shift from 295 K (meV) at $k_{||} = 0$ due to the mechanical compression of the microcavity as a function of temperature resulting in ~3 nm blue-shift with decreasing temperature.

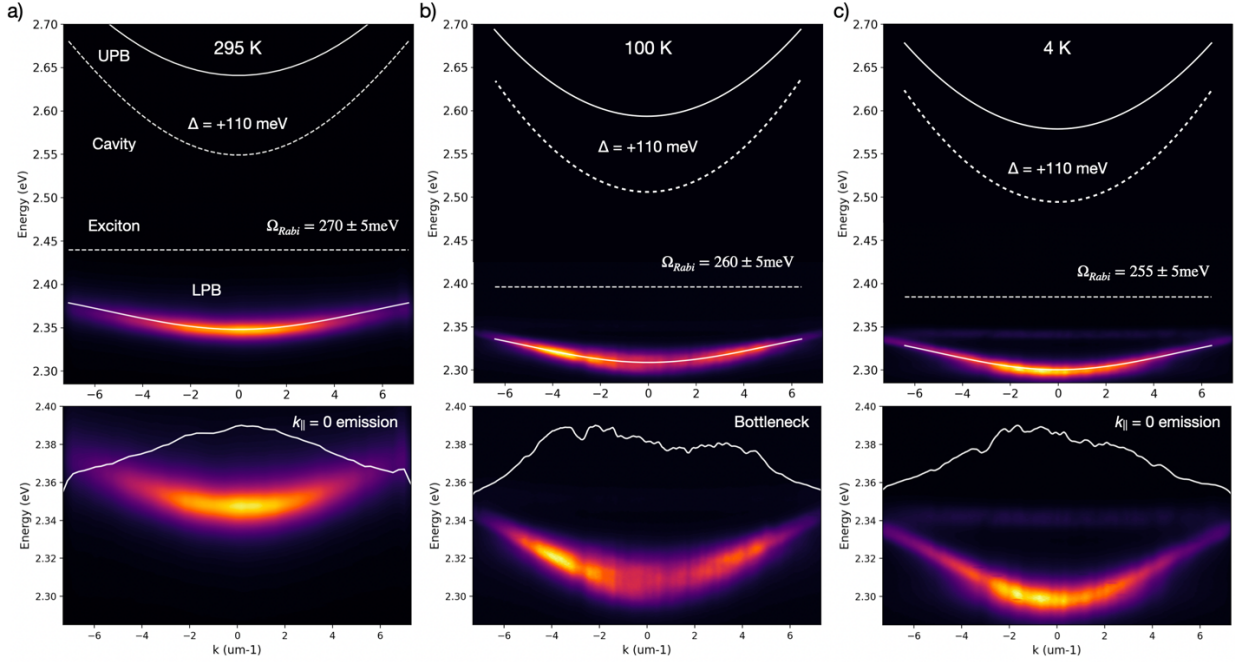

**Figure S10**. Lower polariton branch (LPB) emission for $\hbar\Omega_{Rabi} = 260$ meV fixing the detuning at $\Delta = +110$ meV by selecting a longer cavity length to keep $E_{cav} - E_{exc}$ constant as the exciton energy redshifts with decreasing temperature. Dispersions (upper panels) shown for (a) 295 K, (b) 100 K, and (c) 4 K, revealing the bottleneck effect at intermediate temperatures and emission from $k_\parallel = 0$ at low temperatures (upper and lower polariton branches (solid white traces), exciton energy corresponding to the exciton absorption and bare cavity mode (dashed white traces).[6] Lower panels: LPB photoluminescence (PL) with the energy-integrated PL $k$-space distribution (white trace).

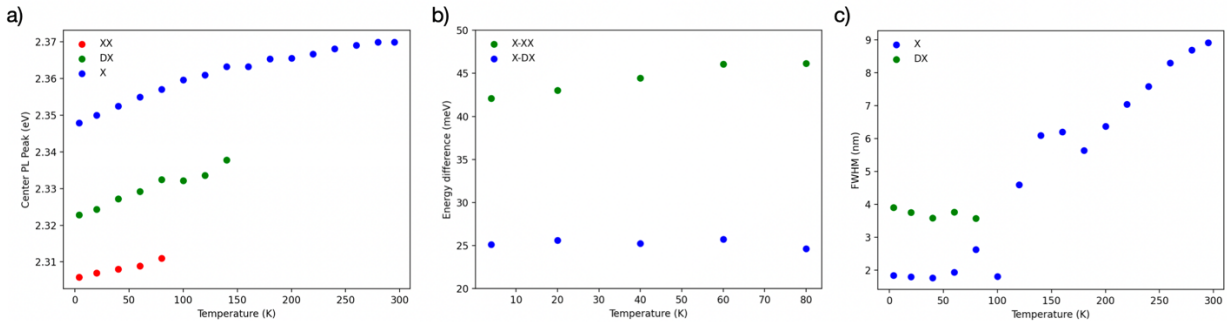

**Figure S11**. (a) The center PL peak (in eV) for the bright exciton (blue), dark exciton (green), and biexciton (red). (b) The difference in PL emission energy between the bright exciton and dark exciton (X-DX, blue) and bright exciton and biexciton (X-XX, green). The full-width half-maximum (FWHM, in nm) for the bright exciton (blue) and dark exciton (green, below 100 K) as a function of temperature, showing a reduction of >4x in the bright exciton FWHM.

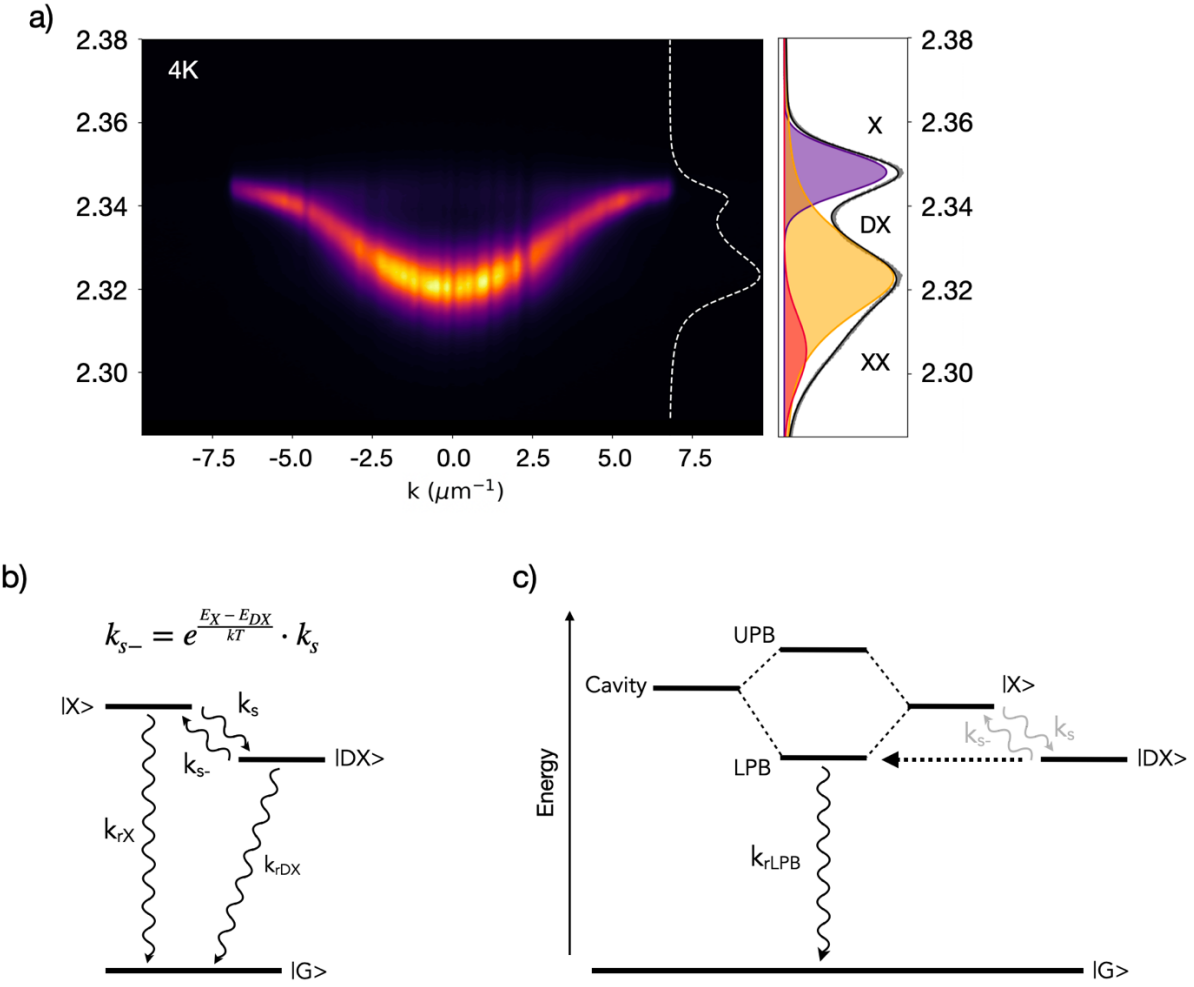

**Figure S12**. (a) 4 K lower polariton branch photoluminescence (PL) ($\hbar\Omega_{Rabi}$ = 175 meV, $\Delta$ = +45 meV) with primary emission from $k_{\parallel}$ = 0. (b) 2D thin film spin-flip process from X to DX with spin-flip rate $k_s$. The spin-flip back to the bright state, $k_{s\text{-}}$, calculated via the Arrhenius relationship as a function of temperature (modified from ref 5).[7] (c) At low-temperature in the microcavity system when $k_s$ is very slow, the strong coupling of the bright exciton to the cavity mode and resulting polariton emission outcompetes the spin-flip from X to DX (gray terms). The DX population generated immediately following photoexcitation benefits from the isoenergetic lower polariton branch mode (dashed arrow from DX to LPB), and can directly and rapidly emit through this resonant mode, taking on the kinetics of the strongly-coupled state.

## Bright/Dark Exciton Dynamics with Photon Recycling

To determine the spin flip rate from the bright to dark state, a set of coupled ODEs was solved to include microscopic reversibility in Eqs. 3-5 and photon recycling. Photon recycling is the ability for a photon emitted following radiative recombination to be waveguided within the film and re-absorbed. The probability of photon escape ($P_{esc}$) is calculated via Eqs. 6-7, in which the indices of refraction for the thin film, substrate, and air interface are taken into account, as well as the optical density (OD) at the wavelengths of emission.11–13 For this system, at 4 K, the PLQE increases 100-fold to ~77%, and the OD increases by nearly a factor of two (Fig. S6, extrapolated).

$$\frac{dn_X}{dt} = -k_{rX}n_X - k_s n_X + e^{(E_X - E_{DX})/k_B T} \cdot k_s n_{DX} + \frac{c}{n_r}\sum_\lambda \alpha_\lambda \gamma_\lambda \tag{3}$$

$$\frac{dn_{DX}}{dt} = -k_{rDX}n_{DX} + k_s n_{DX} - e^{(E_X - E_{DX})/k_B T} \cdot k_s n_{DX} \tag{4}$$

$$\frac{d\gamma_\lambda}{dt} = -\frac{c}{n_r}\sum_\lambda \alpha_\lambda \gamma_\lambda + k_{rX}n_X(1 - P_{esc}) + k_{rDX}n_{DX}(1 - P_{esc}) \tag{5}$$

where $k_{rX}$, $k_{rDX}$, and $k_s$ are the radiative recombination constant for the high energy species, low energy species, and spin-flip rate allowing for interconversion between both species, respectively, $n_X$ and $n_{DX}$ are the high energy and low energy carrier concentrations, respectively, $E_x$ and $E_{DX}$ are the energies of the bright and dark excitons, respectively, $k_B$ is Boltzmann's constant, $c$ is the speed of light, $\alpha_\lambda$ is the absorption coefficient at a given wavelength, $\gamma_\lambda$ is the photon concentration within the film for a given wavelength due to radiative recombination and photon recycling, $n_r$ is the index of refraction, and $P_{esc}$ the probability of a radiatively recombined photon leaving the film within the escape cone.

$$\eta_t = \frac{\Omega_{esc}}{4\pi}T \approx \frac{n_{r2}^3}{n_{r1}(n_{r1} + n_{r2})^2} \tag{6}$$

$$P_{esc} = 10^{-\frac{OD_{PL}}{2}} \cdot \left(n_{t,2D-fs} + n_{t,2D-pmma} + 10^{-OD_{PL}} \cdot \left(n_{t,2D-fs} - n_{t,2D-pmma}\right)\right) \tag{7}$$

where $\eta_t$ is the transmission efficiency, $\Omega_{esc}$ is the solid angle of photon escape, and $n_{rx}$ is the index of refraction of the given material (2D perovskite/fused silica interface and 2D perovskite/PMMA interface). The transmission efficiency for both the 2D perovskite/fused silica interface and the 2D perovskite/PMMA interface is ~17%. The OD of the sister film to the microcavity active layer possessed an OD of ~0.45 at the PL emission wavelength at 4 K. Eq. 7 takes into account the various transmission efficiencies depending on the interface through which a photon escapes the film (in this instance, the interfaces have nearly the same index of refraction contrast). We estimate the probability of photon escape for these highly absorbing 2D thin films to be $P_{esc} = 20\%$.

The impact of photon recycling serves to increase the steady-state carrier density, effectively increasing the average radiative lifetime within the film as compared to a film with no photon recycling.[11] In this system at low temperature, photon recycling effectively feeds the bright state reserve, increasing the pool and allowing a portion of bright excitons to spin-flip into the dark state despite fairly slow spin-flip rates. Based on the spectral overlap between absorption and emission, it is also primarily the radiatively recombined bright excitons that will be re-absorbed and recycled, creating more opportunities to engage in transfer to the energetically lowest-lying dark state at low temperature. We can quantify the number of photon recycling events in the film following the method defined by Pazos-Outón et al. to determine the impact of increasing PLQE with decreasing temperature on photon recycling events in the 2D perovskite film.13

$$Recycling\ events = \frac{1}{1 - f_c \cdot PLQE} \tag{8}$$

$$f_c = 1 - \frac{1}{4n_r^2} \tag{9}$$

where $f_c$ is the confinement factor dictated by the index of refraction, where $n_r = 1.8$. The dependence of the number of recycling events on PLQE is shown in Fig. S13, where, for a PLQE of ~77% (corresponding to 4 K thin film properties), the average number of recycling events is 3.5. In the radiative limit, for PLQE = 100%, the film can sustain 13 recycling events per photon.

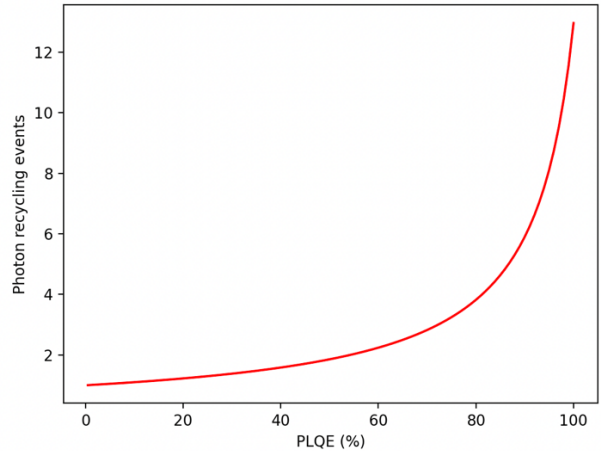

**Figure S13**. Number of photon recycling events in 2D perovskite thin film as a function of film PLQE, indicating that, for PLQE > ~50%, photon recycling events exceed 2 per photon and can contribute significantly to carrier dynamics.[11]

**Bright/Dark Exciton Dynamics without Photon Recycling**

The dynamics can alternatively be simulated excluding the effects of photon recycling (Eq. 10-11). In this way, the bright exciton lifetime is required to exceed 200ps for agreement with the raw data (Fig. S14).

$$\frac{dn_X}{dt} = -k_{rX}n_X - k_s n_X + e^{(E_X-E_{DX})/k_B T} \cdot k_s n_{DX} \quad (10)$$

$$\frac{dn_{DX}}{dt} = -k_{rDX}n_{DX} + k_s n_{DX} - e^{(E_X-E_{DX})/k_B T} \cdot k_s n_{DX} \quad (11)$$

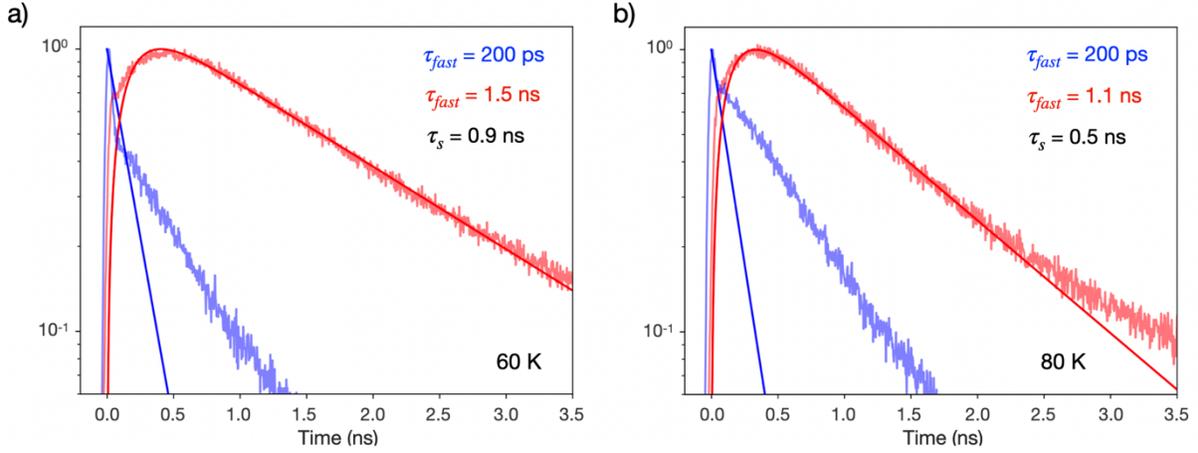

**Figure S14**. Lifetimes of the bright exciton (X) emission (blue) and dark exciton (DX) emission (red) simulated with Eqs. 10-11, excluding photon recycling, at (a) 60 K and (b) 80 K.

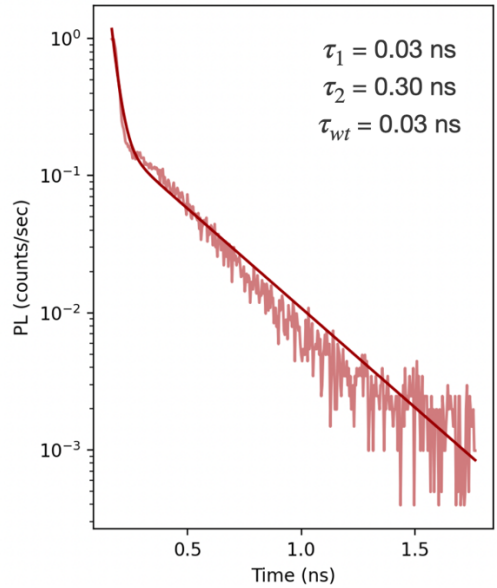

**Figure S15**. Instrument response function of the Toptica wavelength-tunable 80 MHz sub-ps laser and MPD detector. Fits for each trace above ($\tau_1$ = short decay component, $\tau_2$ = long decay component, $\tau_{wt}$ = weighted pulse duration [ns]).

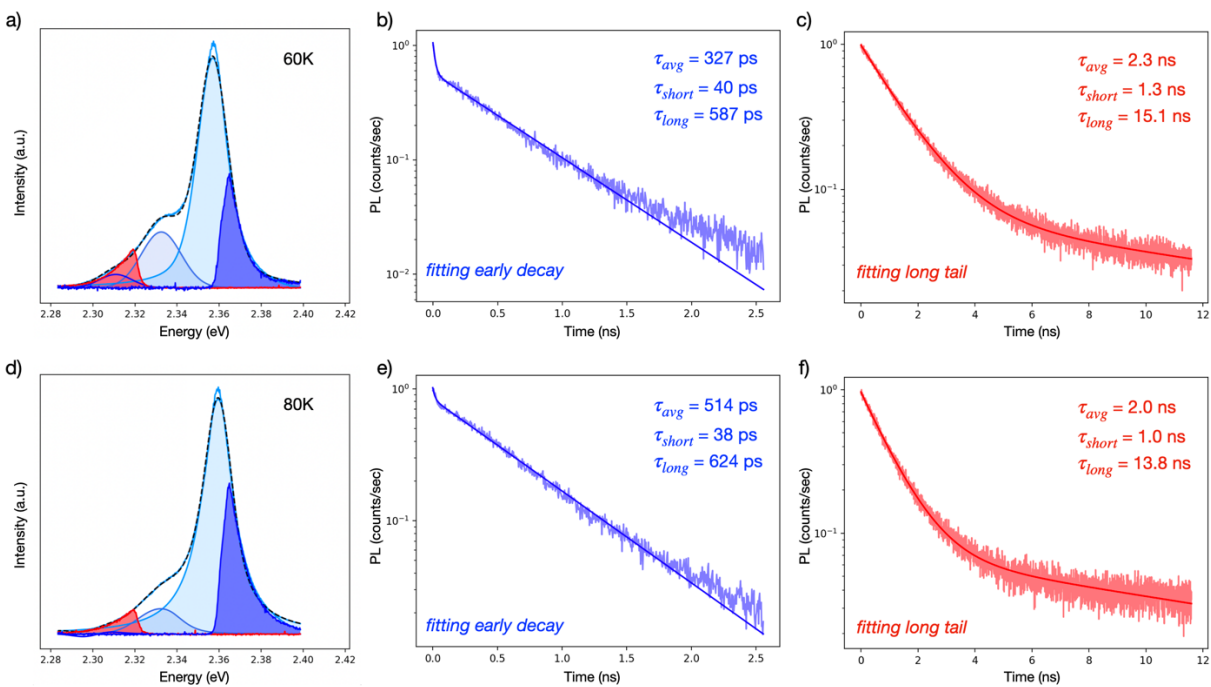

**Figure S16**. (a) 60 K photoluminescence (PL) spectrum of the bare film with multi-peak fitting and spectrally filtered regions highlighted in dark blue/dark red. (b) Time-resolved photoluminescence (TRPL) trace with the tunable edge pass filter (short-pass) showing the short lifetime, fit with an exponential, attributed to the bright exciton (40 ps) and (c) (long-pass) showing only the long tail of the dark exciton (truncating the energy transfer delayed emission portion) to quantify the long component lifetime (15.1 ns). (d-f) PL spectrum at 80 K and extracted high energy short lifetime, fit with an exponential, of 38 ps. The long tail of the dark exciton decreases in lifetime at elevated temperature, with an exponential fit of 13.8 ns.

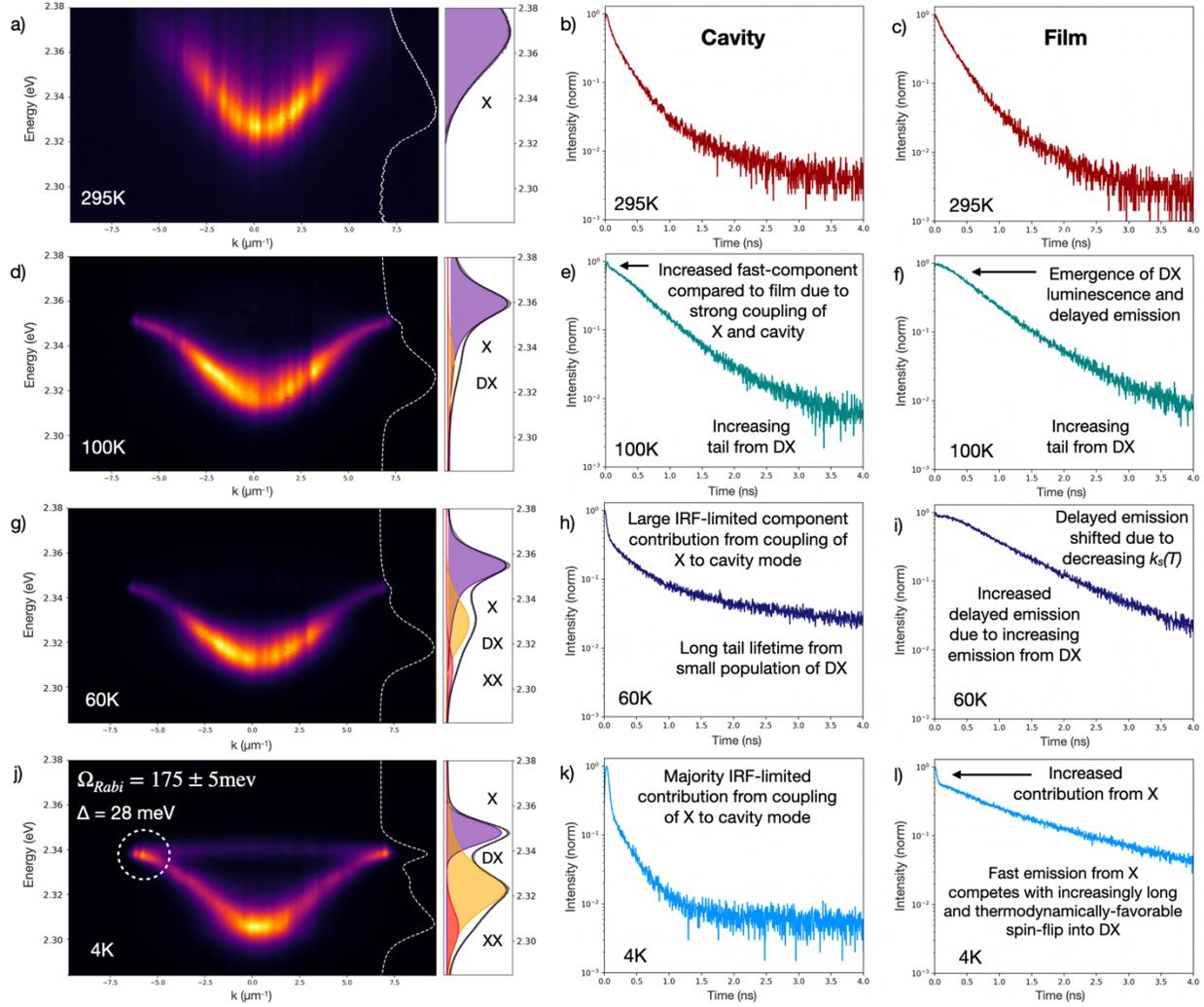

**Figure S17**. The lower polariton branch (LPB) emission from the microcavity ($\hbar\Omega_{Rabi}$ = 175 meV, $\Delta$ = 28 meV, right panel bare 2D film PL spectrum showing bright exciton (X), dark exciton (DX), and biexciton (XX) emission) with time-resolved photoluminescence (TRPL) decay traces for the corresponding LPB cavity emission and bare 2D film at (a,b,c) 295 K, (d,e,f) 100 K, (g,h,i) 60 K, and (j,k,l) 4 K (white dashed circle indicating biexciton-assisted relaxation signature of high $k_{\parallel}$ PL[6]). As temperature decreases, the X emission lifetime decreases and the DX emission emerges with an increasingly long lifetime, visible as a short-timescale fast component with delayed emission into a longer tail ((e,f) green trace, 100 K). In the cavity (e), the extent of delayed emission is reduced as compared to the bare 2D film (f), and the fast component contribution increased due to the additional pathway of coupling the X and cavity mode to form the strongly coupled short-lifetime polariton emissive state competing with the spin-flip from X to DX. For the bare 2D film, further reductions in temperature show (i,l) the X emission contribution increasing at early timescales as its emissive lifetime decreases and the spin-flip rate ($k_s(T)$) slows, with the DX demonstrating an increasingly long emissive lifetime and $k_{s-}(T)$ additionally slowed via the Arrhenius factor (Fig. S13b). Conversely, in the cavity, the IRF-limited strong coupling emission

of X and the cavity mode competes with the spin-flip and begins to dominate the TRPL decay dynamics at 60 K (h), with weak emission contribution to the decay from the DX state resulting in a long lifetime tail. (k) At 4 K in the cavity, the delayed emission due to the slow $k_s(4K)$ and long tail from the DX is not observed, showing nearly exclusively IRF-limited strongly-coupled emission between the X and cavity mode.

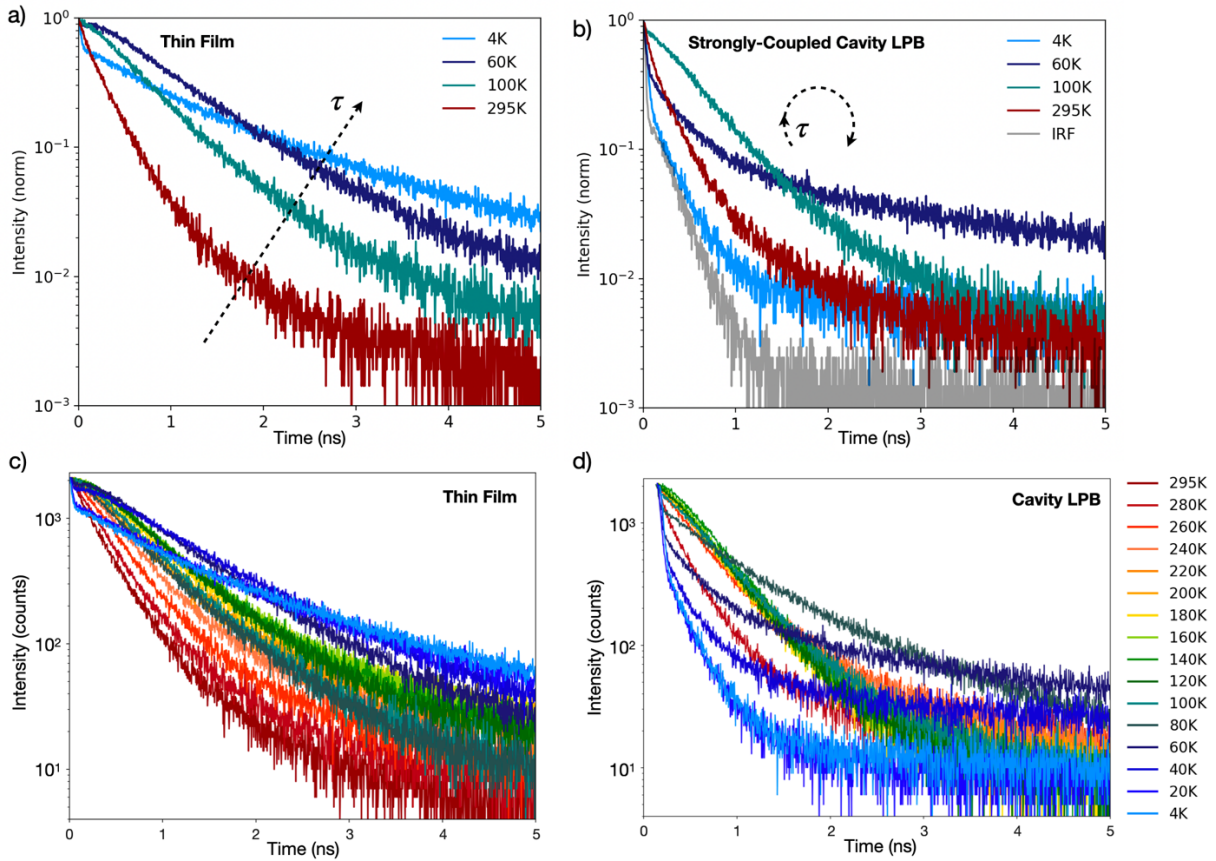

**Figure S18**. (a) The bare 2D perovskite film time-resolved photoluminescence decay (TRPL) traces as a function of temperature show a 295 K lifetime of $\tau_{295K} = 350$ ps, increasing with decreasing temperature. With the emergence of the dark exciton (DX) emission in the PL spectrum, the film lifetime trace develops an initial fast decay component attributed to the bright exciton (X) with a longer lifetime component attributed to the DX. (b) The strongly-coupled microcavity ($\hbar\Omega_{Rabi} = 175$ meV, $\Delta = +28$ meV) lower polariton branch (LPB) emission demonstrates similar trends to the thin film at high temperatures, but deviates sharply at low temperatures, exhibiting only the fast, IRF-limited (40 ps) lifetime of the bare 2D film initial fast decay component. (c) Temperature-dependent 2D thin film TRPL traces with finer temperature steps, showing that, with decreasing temperature, the film lifetime increases to $\tau_{180K} = 740$ ps, consistent with a reduction in non-radiative pathways which quench the lifetime. For temperatures between 180 K and 100 K, we observe delayed emission at early timescales. Below 100 K, with the prominent emergence of the DX, and subsequently XX, emission in the PL spectrum, the film lifetime trace develops an early fast decay component before the flat, delayed emission leading into a long tail. This multi-component lifetime behavior becomes quite pronounced as the system approaches 4 K, and has been observed by Fang *et al.* (light blue trace).[12] (d) The cavity LPB emission with finer temperature steps demonstrates similar trends to the thin film at high temperatures, but deviates sharply at low temperatures, exhibiting only the fast, IRF-limited (40 ps) lifetime of the bare film initial fast decay component attributed to X.

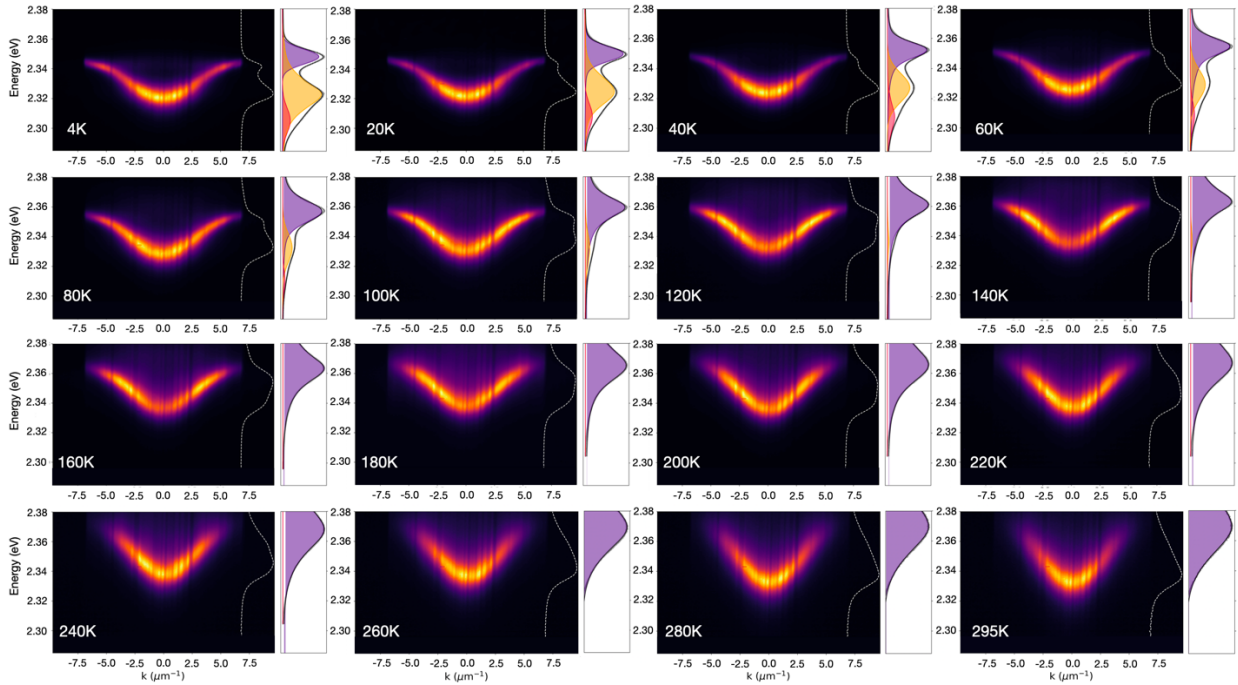

**Figure S19**. Photoluminescence (normalized) k-space temperature series ($\hbar\Omega_{Rabi}$ = 175 meV) from 4 K (upper left) to 295 K (lower right) for $\Delta$ = +45 meV.

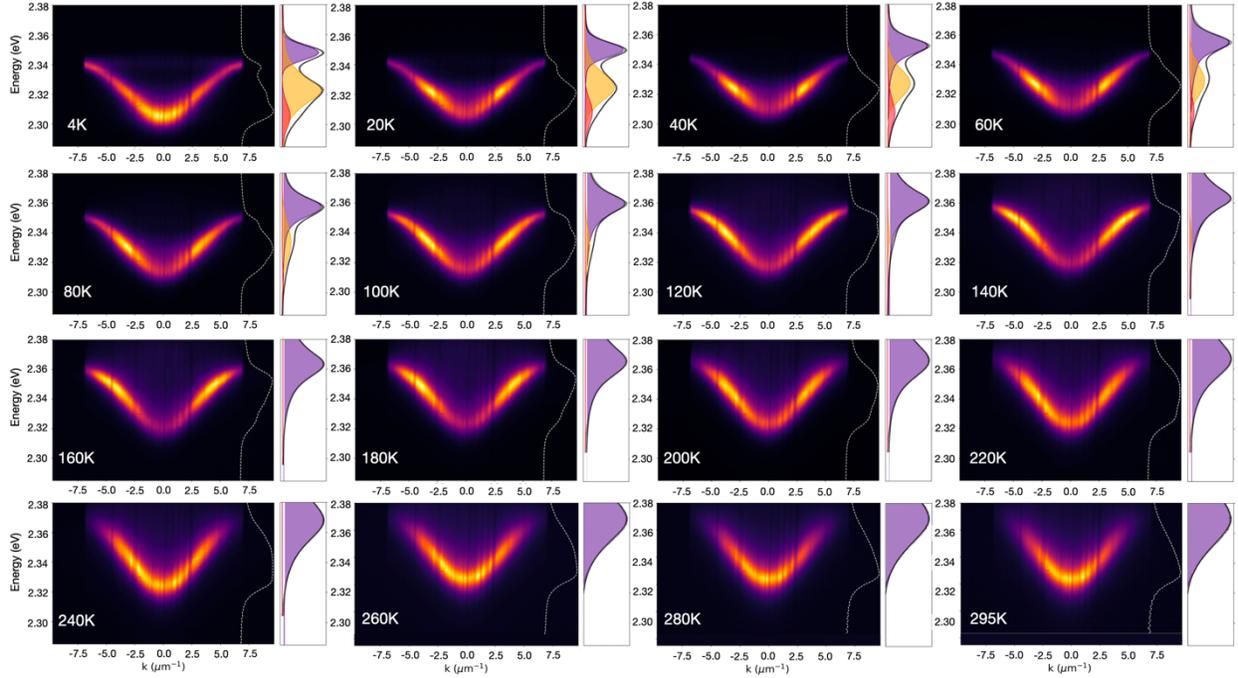

**Figure S20.** Photoluminescence (normalized) k-space temperature series ($\hbar\Omega_{Rabi}$ = 175 meV) from 4 K (upper left) to 295 K (lower right) for $\Delta$ = +28 meV.

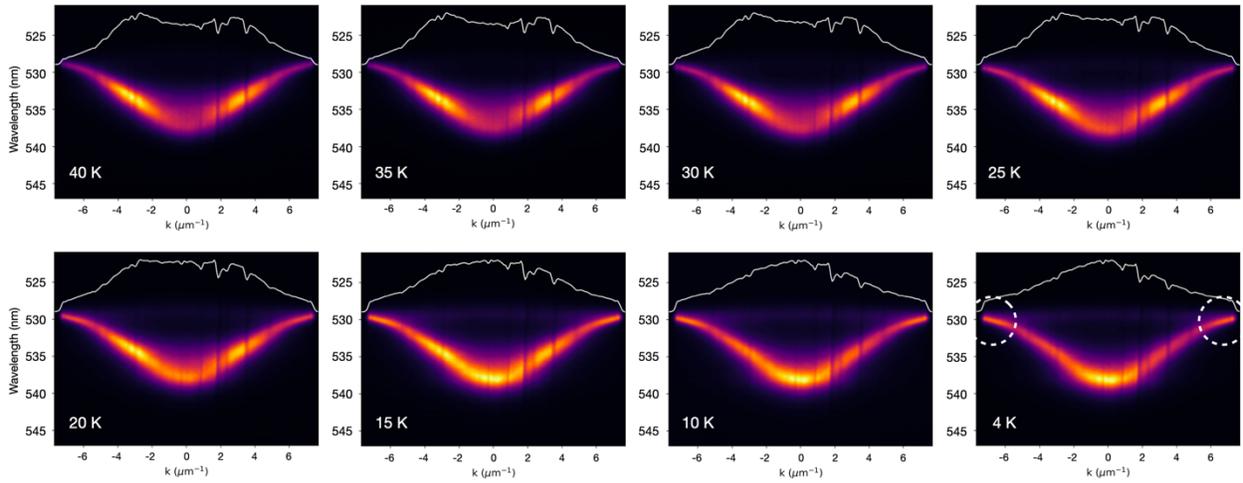

**Figure S21.** Photoluminescence (normalized) k-space temperature series ($\hbar\Omega_{Rabi}$ = 175 meV) from 40 K (upper left) to 4 K (lower right) for $\Delta$ = +28 meV with temperature increments of 5 K to resolve the suppression of the bottleneck and emergence of uncoupled exciton PL and high $k_{||}$ polariton PL from biexciton-assisted relaxation mechanisms.

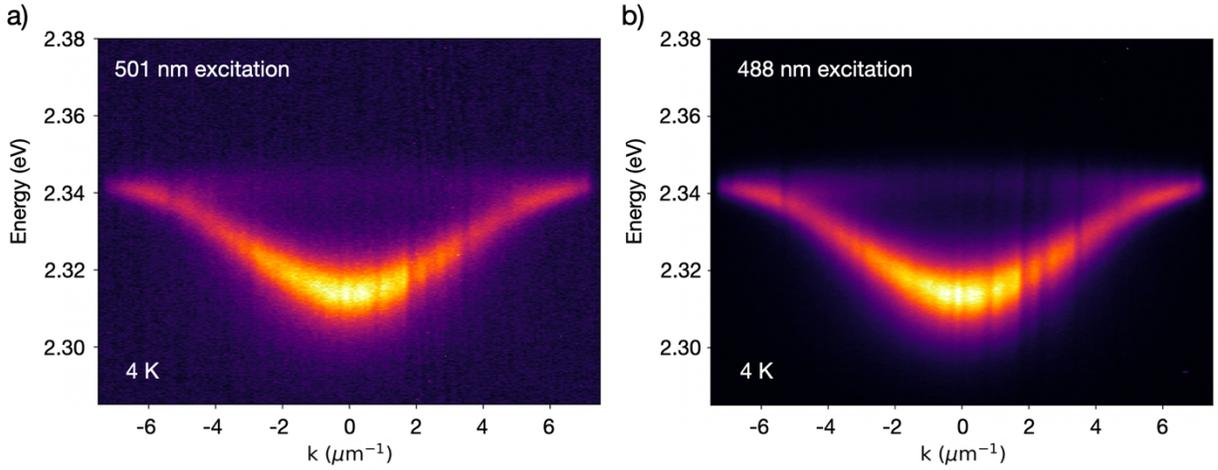

**Figure S22**. 4 K lower polariton branch photoluminescence (PL, normalized) ($\hbar\Omega_{Rabi}$ = 175 meV, $\Delta$ = +35 meV) with 501 nm excitation and 488 nm excitation showing no change to the distribution of PL in k-space as a function of excitation (e.g., via mechanisms such as resonant upper polariton branch excitation). Differences in k-space contrast stem from the decreased absorption cross section at 501 nm as compared to 488 nm.